\documentclass[twocolumn, tight]{aastex631}

\usepackage{xcolor}  

\makeatletter
\newcommand\notsotiny{\@setfontsize\notsotiny{6}{6.5}}
\makeatother

\usepackage{gensymb}
\usepackage{float}
\usepackage{graphicx}
\usepackage{footmisc}

\usepackage{indentfirst}
\usepackage{titletoc}
\usepackage{amssymb}
\usepackage{amsmath}
\usepackage{appendix}
\usepackage{stackengine}
\usepackage{pifont}

\usepackage{color}

\usepackage{soul}
\newcommand\N[1]{{\color{blue} \bf #1}}

\newcommand{\rev}[1]{\textbf{\textcolor{blue}{#1}}}
\renewcommand{\rev}[1]{#1}

\newcommand{\tess}{\textit{TESS}}
\newcommand{\vxm}{SN\,2019vxm}
\newcommand{\atlas}{ATLAS}
\newcommand{\ps}{Pan-STARRS1}
\newcommand{\ztf}{ZTF}
\newcommand{\swift}{\textit{Swift}}
\newcommand{\fermi}{\textit{Fermi}}
\newcommand{\gaia}{\textit{Gaia}}
\newcommand{\lco}{LCOGT}
\newcommand{\sne}{SNe}
\newcommand{\iin}{Type IIn}
\newcommand{\iins}{Type IIns}
\newcommand{\host}{SDSS J195828.83+620824.3}
\newcommand{\mosfit}{\texttt{MOSFiT}}

\begin{document}
\title{SN\,2019vxm: A Shocking Coincidence between \fermi\ and \tess}

\author[0009-0003-8380-4003]{Zachary G. Lane}
\affiliation{School of Physical and Chemical Sciences — Te Kura Matū, University of Canterbury, Private Bag 4800, Christchurch 8140, \\ Aotearoa, New Zealand}

\correspondingauthor{Zachary G. Lane}
\email{zachary.lane@pg.canterbury.ac.nz}

\author[0000-0003-1724-2885]{Ryan Ridden-Harper}
\affiliation{School of Physical and Chemical Sciences — Te Kura Matū, University of Canterbury, Private Bag 4800, Christchurch 8140, \\ Aotearoa, New Zealand}


\author[0000-0002-3825-0553]{Sofia~Rest} 
\affiliation{Department of Computer Science, The Johns Hopkins University, Baltimore, MD 21218, USA}

\author[0000-0002-4410-5387]{Armin~Rest}
\affiliation{William H. Miller III Department of Physics \& Astronomy, Johns Hopkins University,\\ 3400 N Charles St, Baltimore, MD 21218, USA}
\affiliation{Space Telescope Science Institute, 3700 San Martin Drive, Baltimore, MD 21218, USA}

\author[0000-0003-4175-4960]{Conor~L.~Ransome}
\affiliation{Steward Observatory, University of Arizona, 933 North Cherry Avenue, Tucson, AZ 85721, USA}
\affiliation{Center for Astrophysics | Harvard \& Smithsonian, Cambridge, MA 02138, USA}

\author[0000-0001-5233-6989]{Qinan~Wang}
\affiliation{Department of Physics and Kavli Institute for Astrophysics and Space Research, Massachusetts Institute of Technology, \\ Cambridge, MA 02139, USA}

\author[0009-0008-4935-069X]{Clarinda~Montilla}
\affiliation{School of Physical and Chemical Sciences — Te Kura Matū, University of Canterbury, Private Bag 4800, Christchurch 8140, \\ Aotearoa, New Zealand}

\author[0009-0008-8490-0693]{Micaela~Steed}
\affiliation{School of Physical and Chemical Sciences — Te Kura Matū, University of Canterbury, Private Bag 4800, Christchurch 8140, \\ Aotearoa, New Zealand}
\affiliation{Department of Physics, University of Auckland, Private Bag 92019, Auckland, Aotearoa, New Zealand}

\author[0000-0002-8977-1498]{Igor~Andreoni}
\affiliation{Department of Physics and Astronomy, University of North Carolina at Chapel Hill, Chapel Hill, NC 27599-3255, USA}

\author[0000-0003-1997-3649]{Patrick~Armstrong}
\affiliation{The Research School of Astronomy and Astrophysics, Australian National University, ACT 2601, Australia}
\affiliation{E.O. Lawrence Berkeley National Laboratory, 1 Cyclotron Rd., Berkeley, CA, 94720, USA}
\affiliation{The Research School of Astronomy and Astrophysics, The Australian National University, Canberra, ACT 2611, Australia}

\author[0000-0001-6272-5507]{Peter~J.~Brown}
\affiliation{George P. and Cynthia Woods Mitchell Institute for Fundamental Physics and Astronomy, \\ Department of Physics and Astronomy, Texas A\&M University, College Station, TX 77843, USA}

\author[0000-0001-5703-2108]{Jeffrey~Cooke}
\affiliation{Centre for Astrophysics and Supercomputing, Swinburne University of Technology, John St, Hawthorn, VIC 3122, Australia}
\affiliation{ARC Centre of Excellence for Gravitational Wave Discovery (OzGrav), John St, Hawthorn, VIC 3122, Australia}

\author[0000-0003-4263-2228]{David~A.~Coulter}
\affiliation{William H. Miller III Department of Physics \& Astronomy, Johns Hopkins University,\\ 3400 N Charles St, Baltimore, MD 21218, USA}
\affiliation{Space Telescope Science Institute, 3700 San Martin Drive, Baltimore, MD 21218, USA}


\author[0000-0003-2238-1572]{Ori~Fox}
\affiliation{Space Telescope Science Institute, 3700 San Martin Drive, Baltimore, MD 21218, USA}

\author[0009-0006-7990-0547]{James~Freeburn}
\affiliation{Department of Physics and Astronomy, University of North Carolina, Chapel Hill, NC 27599, USA}
\affiliation{Sydney Institute for Astronomy, School of Physics, The University of Sydney, NSW 2006, Australia}

\author[0000-0003-2783-3603]{Marco~Galoppo}
\affiliation{School of Physical and Chemical Sciences — Te Kura Matū, University of Canterbury, Private Bag 4800, Christchurch 8140, \\ Aotearoa, New Zealand}

\author[0000-0002-3653-5598]{Avishay Gal-Yam}
\affiliation{Department of Particle Physics and Astrophysics, Weizmann Institute of Science, 76100 Rehovot, Israel}

\author[0000-0003-1012-3031]{Jared~A.~Goldberg}
\affiliation{Department of Physics and Astronomy, Michigan State University, East Lansing, MI 48824, USA}
\affiliation{Center for Computational Astrophysics, Flatiron Institute, 162 5th Avenue, New York, NY 10010, USA}


\author[0009-0002-6751-2695]{Christopher~Harvey-Hawes}
\affiliation{School of Physical and Chemical Sciences — Te Kura Matū, University of Canterbury, Private Bag 4800, Christchurch 8140, \\ Aotearoa, New Zealand}

\author[0000-0002-1125-9187]{Daichi~Hiramatsu}
\affiliation{Department of Astronomy, University of Florida, 211 Bryant Space Science Center, Gainesville, FL 32611-2055 USA}

\author[0000-0002-0476-4206]{Rebekah~Hounsell}
\affiliation{University of Maryland Baltimore County, 1000 Hilltop Cir, Baltimore, MD 21250, USA}
\affiliation{NASA Goddard Space Flight Center, 8800 Greenbelt Road, Greenbelt, MD 20771, USA}

\author[0009-0008-6765-5171]{Brayden~Leicester}
\affiliation{School of Physical and Chemical Sciences — Te Kura Matū, University of Canterbury, Private Bag 4800, Christchurch 8140, \\ Aotearoa, New Zealand}

\author[0009-0007-3760-515X]{Kl\'ara Lelkes}
\affiliation{HUN-REN CSFK, Konkoly Observatory, MTA Centre of Excellence, Konkoly Thege Mikl\'os \'ut 15-17, Budapest, 1121 Hungary}
\affiliation{ELTE E\"otv\"os Lor\'and University, Institute of Physics and Astronomy, P\'azm\'any P\'eter s\'et\'any 1, Budapest, Hungary}

\author[0000-0002-8304-1988]{Itai Linial}
\affiliation{Columbia Astrophysics Laboratory, Columbia University, New York, NY 10027, USA}

\author[0000-0002-8159-1599]{L\'aszl\'o Moln\'ar}
\affiliation{HUN-REN CSFK, Konkoly Observatory, MTA Centre of Excellence, Konkoly Thege Mikl\'os \'ut 15-17, Budapest, 1121 Hungary}
\affiliation{ELTE E\"otv\"os Lor\'and University, Institute of Physics and Astronomy, P\'azm\'any P\'eter s\'et\'any 1, Budapest, Hungary}

\author[0000-0001-8385-3727]{Thomas~Moore}
\affiliation{Space Telescope Science Institute, 3700 San Martin Drive, Baltimore, MD 21218, USA}

\author[0000-0001-8078-6901]{Pierre~Mourier}
\affiliation{School of Physical and Chemical Sciences — Te Kura Matū, University of Canterbury, Private Bag 4800, Christchurch 8140, \\ Aotearoa, New Zealand}


\author[0000-0002-2028-9329]{Anya~E.~Nugent}
\affiliation{Center for Astrophysics | Harvard \& Smithsonian, Cambridge, MA 02138, USA}

\author[0009-0001-1554-1868]{David O'Neill}
\affiliation{School of Physics and Astronomy, University of Birmingham, Birmingham B15 2TT, UK}
\affiliation{Institute for Gravitational Wave Astronomy, University of Birmingham, Birmingham B15 2TT, UK}


\author[0009-0001-6992-0898]{Hugh~Roxburgh}
\affiliation{School of Physical and Chemical Sciences — Te Kura Matū, University of Canterbury, Private Bag 4800, Christchurch 8140, \\ Aotearoa, New Zealand}
\affiliation{International Centre for Radio Astronomy Research, Curtin University, Bentley, WA 6102, Australia}


\author[0000-0002-2798-2943]{Koji Shukawa}
\affiliation{William H. Miller III Department of Physics \& Astronomy, Johns Hopkins University,\\ 3400 N Charles St, Baltimore, MD 21218, USA}

\author[0000-0002-8229-1731]{Stephen~J.~Smartt}
\affiliation{Astrophysics Research Centre, School of Mathematics and Physics, Queen’s University Belfast, Belfast BT7 1NN, UK}
\affiliation{Astrophysics sub-Department, Department of Physics, University of Oxford, Keble Road, Oxford, OX1 3RH, UK}

\author[0000-0001-5510-2424]{Nathan~Smith}
\affiliation{Steward Observatory, University of Arizona, 933 North Cherry Avenue, Tucson, AZ 85721, USA}

\author[0000-0001-9535-3199]{Ken~W.~Smith}
\affiliation{Astrophysics Research Centre, School of Mathematics and Physics, Queen’s University Belfast, Belfast BT7 1NN, UK}
\affiliation{Astrophysics sub-Department, Department of Physics, University of Oxford, Keble Road, Oxford, OX1 3RH, UK}


\author[0009-0007-0282-7422]{Sebastian~Vergara~Carrasco}
\affiliation{School of Physical and Chemical Sciences — Te Kura Matū, University of Canterbury, Private Bag 4800, Christchurch 8140, \\ Aotearoa, New Zealand}

\author[0000-0002-5814-4061]{V.~Ashley~Villar}
\affiliation{Center for Astrophysics | Harvard \& Smithsonian, Cambridge, MA 02138, USA}
\affiliation{The NSF AI Institute for Artificial Intelligence and Fundamental Interactions}

\author[0000-0001-8764-7832]{J\'ozsef Vink\'o} 
\affiliation{HUN-REN CSFK, Konkoly Observatory, MTA Centre of Excellence, Konkoly Thege Mikl\'os \'ut 15-17, Budapest, 1121 Hungary}
\affiliation{ELTE E\"otv\"os Lor\'and University, Institute of Physics and Astronomy, P\'azm\'any P\'eter s\'et\'any 1, Budapest, Hungary}
\affiliation{Department of Experimental Physics, University of Szeged, D\'m t\'er 9, Szeged, 6720 Hungary}

\author[0009-0005-7414-3965]{Tal~Wasserman}
\affiliation{Department of Particle Physics and Astrophysics, Weizmann Institute of Science, 76100 Rehovot, Israel}


\author[0000-0002-0632-8897]{Yossef~Zenati}
\affiliation{William H. Miller III Department of Physics \& Astronomy, Johns Hopkins University,\\ 3400 N Charles St, Baltimore, MD 21218, USA}
\affiliation{Department of Natural Sciences, The Open University of Israel, Ra’anana 4353701, Israel}
\affiliation{Astrophysics Research Center of the Open University (ARCO), Ra’anana 4353701, Israel}

\author[0000-0001-8985-2493]{Erez~A.~Zimmerman}
\affiliation{Department of Particle Physics and Astrophysics, Weizmann Institute of Science, 76100 Rehovot, Israel}











\begin{abstract}

Shock breakout and, in some cases, jet-driven high-energy emission are increasingly recognized as key signatures of the earliest phases of core-collapse supernovae, especially in \iin\ systems due to their dense, interaction-dominated circumstellar environments. We present a comprehensive photometric analysis of \vxm, a long-duration, luminous Type IIn supernova, $M_V^{}=-21.41\pm0.05\;{\rm mag}$, observed from X-ray to near-infrared. \vxm\ is the first superluminous supernovae \iin\ to be caught with well-sampled \tess\ photometric data on the rise and has a convincing coincident X-ray source at the time of first light. The high-cadence \tess\ light curve captures the early-time rise, which is well described by a broken power law with an index of $n=1.41\pm0.04$, significantly shallower than the canonical $n=2$ behavior. From this, we constrain the time of first light to within $7.2\;\rm hours$. We identify a spatial and temporal coincidence between \vxm\ and the \rev{hard} X-ray/\rev{gamma-ray} transient \rev{GRB\;191117A}, corresponding to a $3.3\sigma$ association confidence. Both the short-duration X-ray event and the lightcurve modeling are consistent with shock breakout into a dense, asymmetric circumstellar medium, indicative of a massive, compact progenitor such as a luminous blue variable transitioning to Wolf-Rayet phase embedded in a clumpy, asymmetric environment.

\end{abstract}



\section{Introduction} \label{sec:intro}

\iin\ are a hydrogen-rich subclass of Supernovae (\sne) that are characterized by their narrow spectral features (often with velocity dispersions $\sim$100$\;{\rm km\,s^{-1}}$) and the presence of hydrogen in the spectrum \citep{Arcavi_2017}. Unlike most other classifications of \sne, \iins\ are not associated with a specific explosion mechanism or progenitor system but instead significant interaction with the circumstellar medium (CSM) in the surrounding environment \citep{Schlegel_1990}, typically created by episodic or eruptive mass loss \citep{Moriya_2023}. In many cases, it remains uncertain whether the progenitor star was stripped before interacting with a hydrogen-rich CSM --- as seen in SN Ia-CSM events --- or whether it retained its hydrogen envelope, as indicated by Type IIn supernovae exhibiting broad hydrogen lines originating from the ejecta \citep{Sharma_2023}. \rev{Events such as SN\,2002ic \citep{Hamuy_2003_ic, Chugai_2004} and SN\,2005gj \citep{Aldering_2006} provide some of the first potential evidence for the SN Ia-CSM subclass.}

Due to the high levels of CSM interaction that seemingly overpowers other sources such as radioactive ${}^{56}$Ni decay, it is expected that the progenitor system undergoes significant mass loss prior to the time of explosion. Because a high mass-loss rate $\gtrsim 10^{-3}\,M_\odot^{}/{\rm yr}$ is required for wind breakouts \citep{Drout_2014, Gezari_2015, Ofek_2014_interaction, Waxman_2017}, massive stellar models that undergo \rev{tumultuous} mass-loss changes and have unstable envelopes such as Red, Yellow, and Blue Supergiants (RSGs, YSGs, BSGs), are likely to be possible progenitors \citep{Arcavi_2017, Smith_2017_book}. There has been direct evidence (e.g., SN\,2005gl \citep{Gal-Yam_2007, Gal-Yam_2009}; SN\,2009ip \citep{Smith_2010,Mauerhan_2013}) that Luminous Blue Variables (LBVs; \citealt{Smith_2026}), which undergo turbulent histories and mass-loss in the transition to the Wolf-Rayet phase, are a progenitor type for \iin\ \sne. In addition, about 50\% of all \iin\ \sne\ experience observable ``precursor'' events within approximately four months before core-collapse, releasing significant amounts of mass into the environment \citep{Smith_2010, Mauerhan_2013, Ofek_2014_precursor, Arcavi_2017, Strotjohann_2021}. 

Another leading theoretical mechanism for the formation of CSM around \iin\ \sne\ is binary interaction \citep{Smith_Arnett_2014, Smith_2017_book}. The binary interaction scenario offers several advantages as it can naturally account for asymmetric or non-spherical CSM distributions \citep{Smith_2017_book}. However, the wind velocities associated with Roche-lobe overflow in binary systems are typically insufficient to explain the most luminous events, such as SN\,2006gy, SN\,2006tf, and SN\,2010jl \citep{Smith_2014, Smith_2017_book}. Recent studies suggest that binary mass transfer can reproduce the observed brightness distribution of most \iins\ \citep{Ercolino_2025}, though the most luminous \iins\ likely represent a more complex subset. Regardless, binary interaction may remain a key factor in shaping the geometry, density, and overall distribution of the CSM in both steady-wind and eruptive mass-loss environments \citep{Smith_2017_book}.

Due to the diverse progenitor channels and complex interactions with the CSM, \iin\ \sne\ exhibit the broadest range of absolute brightness among all SN types, spanning over six magnitudes \citep{Smith_2017_book}. \iin\ \sne\ are frequently among the most luminous and superluminous\footnote{\rev{There is ongoing debate regarding both the definition and the validity of classifying superluminous \iin\ \sne, as they are not clearly distinct from the broader \iin\ population \citep{Hiramatsu_2024}. Here, we adopt this designation solely to denote their high luminosity, without implying a physically distinct subclass.}} supernovae \citep[SLSNe;][]{Howell_2017}, defined as those with peak magnitudes $M_x \leq -21$ in any optical band \citep{Gal-Yam_2012, Gal-Yam_2019}. Similar to conventional supernovae (SNe), SLSNe are classified by hydrogen content: SLSNe-I lack hydrogen in the spectra, and SLSNe-II are hydrogen rich. While the explosion/progenitor systems are still unknown, most SLSNe-II have the time evolving narrow emission lines reminiscent of \iin\ spectra \citep{Quimby_2018}, suggesting they may represent a luminous extension of the \iin\ population, which we will refer to as SLSNe-IIn. Whether SLSNe-IIn are sufficiently distinct from the broader diversity of \iin\ \sne\ to be considered a separate classification remains under discussion. For clarity, we adopt the superluminous designation to distinguish the more luminous events with which \vxm\ is more closely associated.

Some of these more luminous and superluminous \iin\ include SN\,2003ma \citep{Rest_2011}, SN\,2006gy \citep{Smith_2007, Ofek_2007, Smith_2010_06gy}, SN\,2006tf \citep{Smith_2008}, SN\,2008am \citep{Chatzopoulos_2011}, SN\,2010jl \citep{Smith_2011, Ofek_2014_xray, Ofek_2019}, SN\,2015da \citep{Smith_2024_15da}, and SN\,2017hcc \citep{Smith_2020_hcc, Chandra_2022, Moran_2023}. As the shock traverses the extended CSM structure, SLSNe-IIn can remain bright for years \citep{Smith_2017_book}. The complexity of their CSM interactions makes studying \iins\ --- particularly the most luminous --- productive for understanding star formation and dust production \citep[e.g.][]{Shahbandeh_2024}.

A key feature of core-collapse supernovae, including many \iin\ \sne\ and SLSNe-IIn, is the emergence of early radiation from the stellar surface following explosion. After a core-collapse, the first photons escape from the optically thick stellar surface in a process known as a `shock-breakout' \citep{Waxman_2017}. This occurs when a shock wave, generated deep within the optically thick interior of the progenitor's stellar interior, travels outward and reaches the stellar surface. Upon breakout, a brief but intense flash of X-ray/UV radiation is released, as potentially first seen in \citet{Soderberg_2008}, with a duration ranging from a few seconds to several hours, depending on the progenitor's size, structure, and the shock's energy and geometry \citep{Katz_2010, Matzner_2013, Waxman_2017, Irwin_2021, Irwin_2025}. As the shock-driven outermost stellar layers expand and cool, ejected by the internal shock, they emit UV/optical radiation, typically lasting several days \citep{Chatzopoulos_2012, Waxman_2017}. This early emission often constitutes the first detectable light from a SN \citep{Waxman_2017}. Optical counterparts to these events can be regularly detected for Type II \sne\ \citep{Irani_2024}, with examples including the high-cadence \textit{Kepler} observations of the Type IIP SN KSN\,2011a/d \citep{Garnavich_2016} and the rapidly rising Type IIb SN\,2016gkg \citep{Bersten_2018}. 

In this paper, we analyze the photometric data for \vxm, a \iin\ SN or Superluminous-IIn (SLSN-IIn) -- discovered by the ASAS-SN team \citep{Kochanek_2017} and given the designation ASASSN-19acc \citep{ASAS-SN_2019} -- and the \rev{gamma-ray burst GRB\;191117A detected by the Fermi Gamma-ray Burst Monitor (GBM)}. Many high-cadence surveys including \tess\ and \atlas\ gathered data, giving \vxm\ a well-constrained rise and one of the most complete photometric coverages for any \iin\ SN. Spectroscopic analysis of \vxm\ will be presented in \citealt{Smith_prep}.

\vxm\ was found at a redshift of $z_{\rm hel}^{}\approx0.019$, peaking at a non-extinction corrected absolute magnitude of $M_V^{}=-20.12\,{\rm mag}$\footnote{Converted to the fiducial spatially-flat $\Lambda$CDM cosmology with $H_0^{}=70\;{\rm km\,s^{-1}Mpc^{-1}}$ and $\Omega_{\rm M0}^{} = 0.3$ that can be assumed throughout the rest of the paper.} \citep{Tsvetkov_2024} and originates from the host-galaxy \host. 

In \S\ref{sec:data} we discuss the photometric data available as well as the processing and reduction. In \S\ref{sec:analysis}, we analyze the photometric light curves and data of \vxm\ alongside other \iin\ \sne\ to identify potential similarities and differences. Section \ref{sec:shocks} presents the search for and interpretation of optical and X-ray shock breakouts, including a coincidence between the \fermi\ event \rev{GRB\;191117A} and \vxm. In \S\ref{sec:models} we model the lightcurve of \vxm\ and the host and place constraints on its rise parameters. The main conclusions and discussion are given in \S\ref{sec:conclusion}.

\section{Data \& Reduction} \label{sec:data}



\subsection{TESS} \label{sec:tess}

While the Transiting Exoplanet Survey Satellite (\tess) space telescope mission \citep{Ricker_2014} was originally designed with the primary purpose of capturing exoplanet transits, the high cadence data and wide field of view allow for the imaging and discovery of many transient-like (time-domain astronomy) events. \tess\ has four cameras each with a $24\degree \times 24\degree$ field of view (for a total area of $24\degree \times 96\degree$), imaging a \textit{sector} of the sky \rev{in a single broadband red filter, \tess\textit{-R},} for $\sim$27 days\footnote{Depending on the ecliptic latitude, some targets can appear in multiple sectors consecutively.} \citep{Ricker_2014}. 

For the original mission (April 2018 -- July 2020), each sector was imaged at a cadence of 30~minutes. 
The high temporal cadence and the relatively long observing strategy provides a unique opportunity for the classification and discovery of many unique transient-like phenomena. \tess\ has effectively been used to classify, and even discover many time-evolving high-energy astrophysical phenomena such as supernovae \citep{Fausnaugh_2023_sne, Wang_2023, Wang_2024}, and Gamma-Ray Burst (GRB) afterglows \citep{Fausnaugh_2023_grb, Roxburgh_2024, Jayaraman_2024, Perley_2025}. 

While \tess\ has great temporal resolution, it has relatively poor spatial resolution, where each pixel is approximately $21\arcsec \times 21\arcsec$, with an effective Point-Spread Function (ePSF)\footnote{While it is not uncommon to refer to the ePSF as the Pixel Response Function (PRF) for \tess, we refer to the PRF as the ePSF to be in line with difference-imaging pipelines and methods across various surveys.} with a Full-Width at Half-Maximum (FWHM) typically between $21\arcsec$ and $42\arcsec$ \citep[one to two pixels;][]{tessprf_2022}. Because of the undersampled ePSF, any PSF photometry can be quite dominated by the individual response/sensitivity of the pixels \citep{tessprf_2022}.

The Full Frame Images (FFIs), which are $(2048 \times 2048)\;{\rm pixels}^2$ for each of the 16 CCDs that can be accessed through the Barbara A.~Mikulski Archive for Space Telescopes \rev{\citep[MAST;][]{MAST_2022}}, have undergone ``calibration'', but no subsequent background subtraction. Removing the scattered light background across the detector is highly non-trivial as it evolves rapidly spatially and temporally. The highly elongated orbit between the moon and the Earth causes scattered light from both the Earth and the Moon to enter \tess's optical path, requiring further reduction pipelines to analyze objects \citep{Ridden_2021}. 

For the analysis of \vxm, we utilize the \texttt{TESSreduce}\footnote{\url{https://github.com/CheerfulUser/TESSreduce}} \citep{Ridden_2021} difference-image reduction pipeline which has been successfully used in various astrophysical transient analyses (e.g., \cite{Tinyanont_2022, Wang_2023, Wang_2024, Roxburgh_2024}); and is a major component of \texttt{TESSELLATE},\footnote{\url{https://github.com/rhoxu/TESSELLATE}} the first large-scale transient detection pipeline for \tess\ \citep{Roxburgh_2025}. \texttt{TESSreduce} uses the \texttt{python} package \texttt{calibrimbore}\footnote{\url{https://https://github.com/CheerfulUser/calibrimbore}} to photometrically calibrate the \rev{\tess\textit{-R} filter} to the \ps\ standardized filters.

The \texttt{TESSreduce} package provides a choice between aperture and PSF photometry, where the preferred method is circumstantial. Since neither method is consistently preferred in \tess, we compare both approaches, where the preferred practice should be the method that minimizes the flux scatter and instrumental trends. For our analysis, we find that for \vxm, PSF photometry is the preferred method \rev{as it substantially reduces light contamination from other sources}, and therefore, we use the PSF reduction \rev{on the \tess\ difference-images}.

\begin{figure}
    \centering
    \includegraphics[width=1\linewidth]{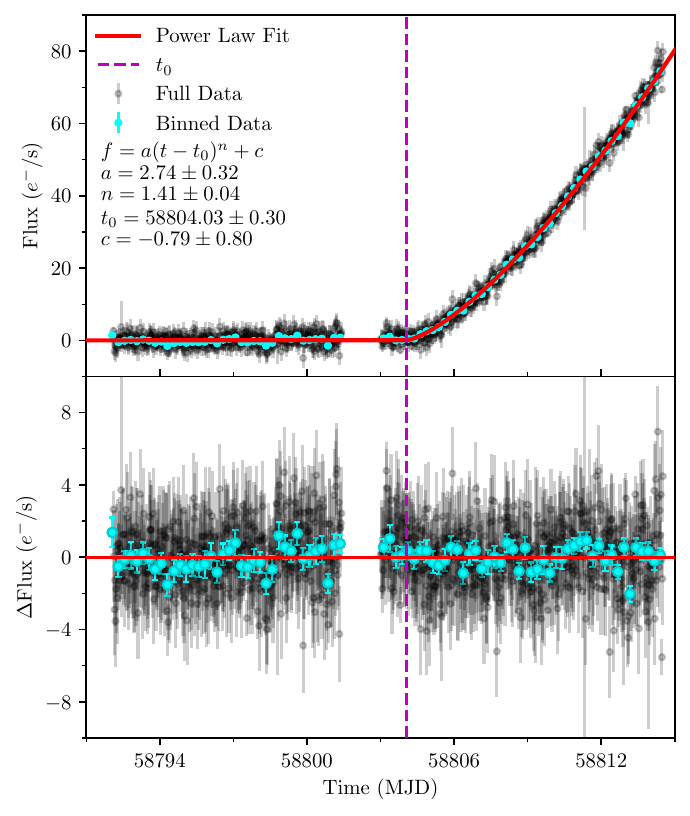}
    \caption{The rise of \vxm\ in the native 30~min cadence and 6~hr bins after adjusting the baseline to follow zero-flux. Both plots show the complete rise, and the data binned in $6\;{\rm hr}$ sections. \textbf{Top:} the flux for the Sector~18 with a simple broken power-law fit to the rise. We show the value for $c$ despite the baseline correction already having been applied in the figure. \textbf{Bottom:} the residual, $\Delta {\rm Flux} = {\rm Flux}_{\rm l.c.}^{} - {\rm Flux}_{\rm p.law.}^{}$, between the lightcurve and the fit power-law, to test for significant deviations in the lightcurve.}
    \label{fig:tess_resid}
\end{figure}

Approximately midway through a sector, \tess\ undergoes \textit{downlinking}, where the telescope can not be used for observations for $\sim$16~hrs while the data onboard the craft gets sent to Earth. The downlinking creates two distinct regions, each with minor differences in calibration\rev{, typically a constant offset correction to the baseline of approximately $0.1\to5\;\rm counts$}. For exoplanetary science this is dealt with by normalizing all of the lightcurves, but for difference-imaging transients this can be complex. To ensure the baseline for the two regions is identical we fit a simple broken power-law to the rise,
\begin{equation}\label{eq:powerlaw}
    f = \Theta(t-t_0) \cdot \left[ a(t-t_0)^{n} + c \right] \; ,
\end{equation}
where $\Theta(t-t_0)$ is the Heaviside function acting at the time of first light $t_0$, $a$ is a flux scaling factor, $n$ is a power-law index, and $c$ is a constant linear vertical shift. The modeled $t_0$ is relative to the time of first light for the specific instrument, where different telescopes with different sensitivities may infer an earlier $t_0$. A broken power-law is the simplest model that can be used to explain the early-rise, as seen in \autoref{fig:tess_resid}. While other work assumes the rise is $\propto t^2$ \citep{Nyholm_2020}, we consider the high-cadence \tess\ data and the rise color-dependence to constrain $n$ which provides a qualitatively greater fit than assuming $n=2$. In the presence of more complex dynamics such as seen in the Type Ia SN\,2023bee \citep{Wang_2024}, or shock breakouts as seen in \textit{Kepler} high-cadence Type IIP SN \citep{Garnavich_2016}, more complicated models would have to be employed. We tested a range of $t_0$ values, selecting the value (and subsequent model) that minimizes the error-weighted $\chi^2/{\rm d.o.f.}$, and applying the baseline corrections to the rise of the supernova. The errors presented for the broken power-law come from the covariance from the model fitting added in quadrature with the bootstrap resampling errors from the weighted sample of the other $t_0$ cuts. We note that the modeled $t_0$ is relative to the sensitivity of the \tess\ telescope and it is possible that larger telescopes would find an earlier rise.

To \rev{flux} calibrate the \tess\ \rev{data}, we use \atlas\ magnitudes taken about $0.5\,{\rm days}$ \rev{before} the first spectra available on the \texttt{Transient Name Server} (TNS)\footnote{\url{https://www.wis-tns.org/}\label{foot:tns}}. \rev{We assume that (i) the power-law relation describing the rise remains valid at the epoch of the first spectrum, a few days after the final \tess\ observation, and (ii) \vxm\ does not undergo any significant spectral evolution over a timescale of $0.5\,{\rm days}$. }

\rev{We first extrapolate the best-fit power-law model to the time of the first spectrum to predict the corresponding \atlas\ magnitude, adopting a standard zeropoint \citep{Vanderspek_2018}. The difference between this predicted value and the observed \atlas\ magnitude defines the magnitude offset between the model \tess\ flux and the \atlas\ measurement. We apply this offset ($-0.052\,{\rm mag}$) to the \atlas\ magnitude.\footnote{While we would not expect the magnitude offset to be identical between the two bands, they are a good approximation over such a small magnitude shift.}}

\rev{Using the corrected AB calibrated \atlas\ magnitude}, we performed synthetic photometry with \texttt{pysynphot} \citep{STScI_2013} and the relevant passbands to derive the expected magnitude for the \tess\textit{-R} band, calculating the zeropoint to be ${\rm zp} = 20.57$, which is $0.13\,{\rm mag}$ higher than the prototypical Vega-system zeropoint of 20.44 (based on the Cousins \textit{I}-band \citep{tess_readme_2025}) commonly adopted in \tess\ photometry \citep{Vanderspek_2018}.

\subsection{ATLAS} \label{sec:atlas}

\begin{figure}
    \centering
    \includegraphics[width=\linewidth]{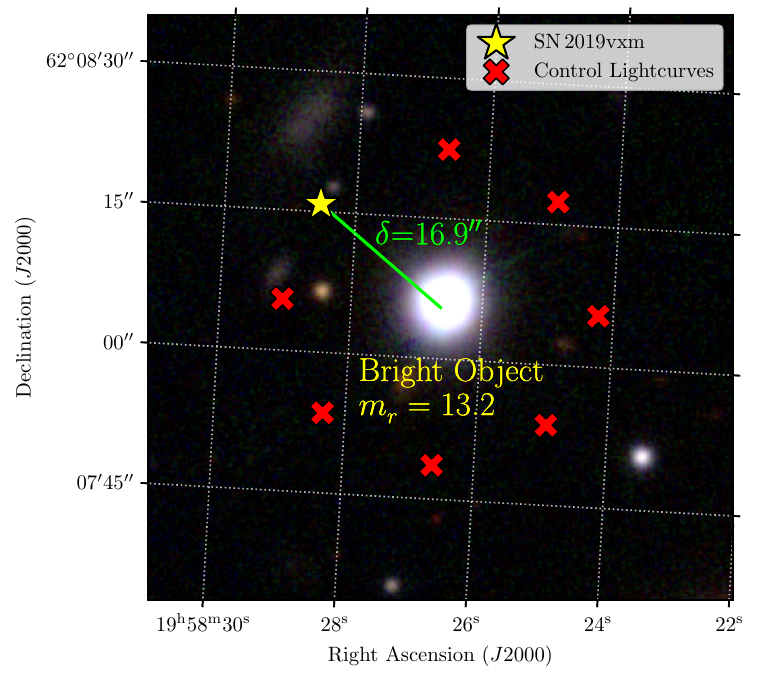}
    \caption{
        An overlay of \vxm\ and the control lightcurve positions used in the
        \texttt{ATClean} reduction on a tricolor (\textit{g}, \textit{r}, \textit{i}) \ps\ image,
        colored using the \texttt{AstroColour}
        package. The control lightcurve positions were chosen relative to the bright source near \vxm,
        as that will be the main source of contamination.
    }\label{fig:atlas_loc}
\end{figure}

The Asteroid Terrestrial-impact Last Alert System (\atlas) program is a series of four Schmidt $0.5\,{\rm m}$ telescopes (two at the time of \vxm), primarily designed for asteroid and Near Earth Objects detection \citep{Tonry_2018_survey}. Despite the designed purpose, \atlas\ has discovered over 4000 supernovae and other transients up to $\sim$19.5$\,{\rm mag}$ \citep{Tonry_2018_survey, Smith_2020_atlas}. \atlas\ primarily uses two filters, \textit{ATLAS-o} (orange) and \textit{ATLAS-c} (cyan),\footnote{\atlas\ does have other bands, including a \textit{t}-band (tomato), however, \vxm\ was not captured in these bands.} that were originally calibrated using \textit{APASS} (in the south) and \ps\ (north of $-30\degree\,{\rm Decl.}$) photometry, but now include \gaia, the Tycho-2 catalog, the Yale Bright Star Catalog, and \textit{SkyMapper} \citep{Tonry_2018_catalogue}. \textit{ATLAS-o} and \textit{ATLAS-c} can be approximated using linear combinations of the \ps\ $g$, $r$, $i$ filters for stellar spectral energy distributions as given by \citet[Eqn. 2]{Tonry_2018_survey}

Unlike \tess\, which observes a section of the sky continuously, \atlas\ is a ground-based program and on a clear night images almost all of the visible sky at four epochs to form quads \citep{Tonry_2018_survey}. Therefore, the nominal cadence is approximately four times every day in the \textit{o}-band, and less in the \textit{c}-band.

We produce the \atlas\ lightcurve using \texttt{ATClean}\footnote{\url{https://github.com/srest2021/atclean}} \citep{Rest_2025}, which is built upon the methods described in \citet{Tonry_2018_survey} and \citet{Smith_2020_atlas} to perform forced PSF photometry on \vxm. \texttt{ATClean} performs multiple forced difference-imaged photometry on the target and at several surrounding non-crowded locations without bright sources, as seen in \autoref{fig:atlas_loc}, to produce \textit{control lightcurves}. This procedure removes outliers by testing if uniform changes in brightness are detected across the control samples. The control lightcurve positions can be seen in \autoref{fig:atlas_loc}.

\subsection{Pan-STARRS1} \label{sec:ps1}

The \ps\ survey is one of the most complete and well-calibrated photometric surveys, covering the entire northern sky (${\rm Dec.}>-30\degree$) since 2007
\citep{Chambers_2016}.  Initially calibrated using synthetic magnitudes and the \textit{Sloan Digital Sky Survey}, it now relies on its own catalog for calibration due to the large data volume \citep{Magnier_2020_photom}. \ps\ employs \textit{Iteratively Weighted Least-Squares} photometric techniques to determine the flux of each target following extensive pixel de-trending, as outlined in \citet{Waters_2020}, which is applied to the output difference-imaged forced photometry AB magnitude lightcurve \citep{Magnier_2020_photom}. Because \vxm\ is in a rarely visited field of the \ps\ survey, only a few detected points exist. There is no detection of any excess flux or precursor activity in 98 images taken between MJD 57505 and 58765. These are mostly 45 sec $i$-band images, with around 10 $z$ or $y-$band 30 second exposures. The 3$\sigma$ forced limits for the $i-$band images are 
$i\gtrsim21.2$. 

\subsection{ZTF} \label{sec:ztf}

The \textit{Zwicky Transient Facility} (\ztf) \citep{Bellm_2019} is a wide-field Northern Hemisphere survey that operates with a cadence of $\sim$2~days. The survey utilizes two telescopes, a $1.2\,{\rm m}$ and a $1.5\,{\rm m}$, at Palomar Observatory, primarily aimed at detecting transient events such as novae and GRBs with a photometric precision of 8 to 25 mmag \citep{Masci_2019}. The data products for \vxm\ were created using the forced photometry service descibed in \citet{Masci_2023}, which computes AB magnitudes using the extensive \ps\ catalog as a reference for computing the zeropoint of the field.

\subsection{Konkoly} \label{sec:konkoly}

\rev{The Konkoly Observatory in Hungary houses several telescopes ranging in size from $0.4\,\rm m$ to $1.0\,\rm m$ at Piszk\'estet\H{o} Mountain Station. The data were taken with the $0.8\,\rm m$ Ritchey-Cr\'etien robotic telescope, in Johnson-Cousins \textit{BV} and Sloan \textit{griz} filters. The data were processed with standard \texttt{IRAF} routines, using aperture photometry. Measurements were standardized using the Pan-STARRS Data Release 1 (PS1) magnitudes of the stars near the supernova in the field of view \citep{Chambers_2016}. For a more detailed description, we refer to \citet{Barna_2023}; a separate analysis of the Konkoly data, together with the photometry by \citet{Tsvetkov_2024}, will be presented by Lelkes et al.\ (in prep.).}  

\subsection{Fermi} \label{sec:fermi}

\begin{figure}
    \centering
    \includegraphics[width=1\linewidth]{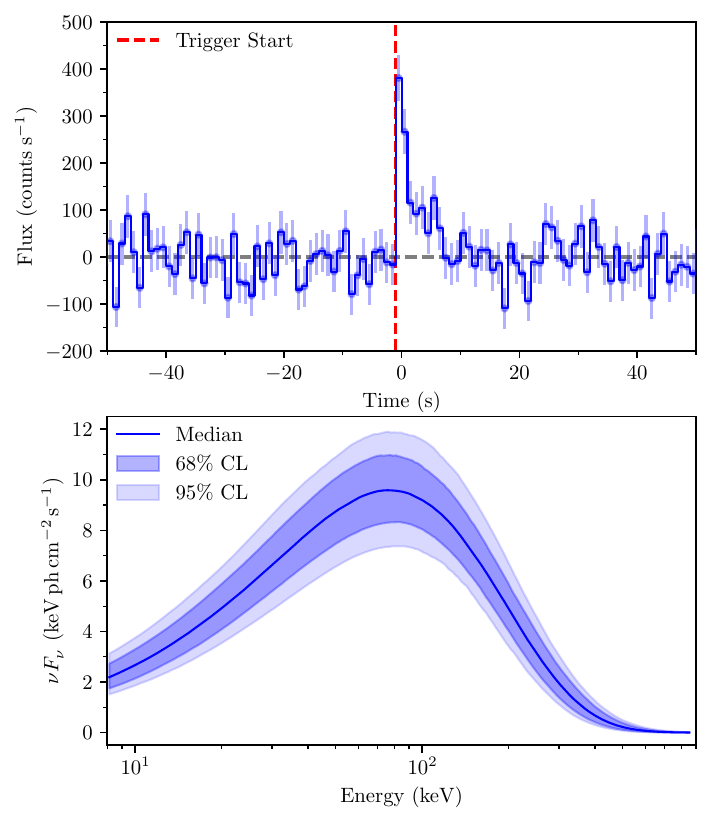}
    \caption{The \rev{summed detectors N6 and N7 X-ray/}gamma-ray \fermi\ GBM data within the energy range 8--900$\;{\rm keV}$ binned in increments of \rev{$1\,{\rm s}$ for the displayed light curve and $0.512\,{\rm s}$ for the spectral analysis}. \textbf{Top}: the background subtracted \fermi\ lightcurve. \textbf{Bottom}: the energy-calibrated\rev{, and forward modeled spectrum using a Comptonized model} showing the peak energies of the X-rays photons detected at the time of the event\rev{. The 68\% and 95\% confidence regions for the Comptonized model are shown by the shaded regions.}}
    \label{fig:fermi_lc}
\end{figure}


The \fermi\ \rev{\textit{Gamma-Ray Space Telescope}} conducts an all-sky survey for photons in the energy range of $\sim$8$\,{\rm keV}$ to $\sim$300$\,{\rm GeV}$\footnote{Corresponding to a wavelength range of $(\sim1.5\times10^{-10}$ to $\sim$4$\times10^{-18})\,{\rm m}$.}. The primary instrument, the \fermi\ Large Area Telescope \citep{Atwood_2009}, has a wide field of view of $\sim7880\,{\rm deg^2}$, covering about 20\% of the sky, and is sensitive to gamma-rays in the range of $\sim$20$\,{\rm MeV}$ to $\sim$300$\,{\rm GeV}$ \citep{Atwood_2009}.

The secondary instrument, the Gamma-ray Burst Monitor (GBM), \rev{observed \rev{GRB\;191117A}, which we associate with \vxm (see \S\ref{sec:loc} for details on the associated probability). The GBM} is equipped with 12 sodium iodide (NaI) scintillators and two bismuth germanate (BGO) scintillators. The detectors are capable of detecting X-rays and low-energy gamma-rays (up to $1\,{\rm MeV}$ for NaI and $40\,{\rm MeV}$ for BGO) over the entire sky not occulted by the Earth \citep{Meegan_2009, Poolakkil_2021}. The Time-Tagged Event (TTE) data from the GBM catalog objects \citep{von_Kienlin_2014, Gruber_2014, Bhat_2016, von_Kienlin_2020} is highly versatile, recording the time and energy of each event with a cadence of $5\,\mu{\rm s}$, often binned for a `cleaner' analysis. With the time and \rev{energy} information, a conventional lightcurve product can be created, while the energy and timing information can create a \rev{forward-folded} spectrum \rev{from} the incoming photons as seen in \autoref{fig:fermi_lc}.


To analyze the GBM data, we use the \texttt{gdt-fermi}\footnote{\url{https://astro-gdt.readthedocs.io/projects/astro-gdt-fermi/en/latest/index.html}} package \citep{Goldstein_2023}, which is built upon the \texttt{gdt-core} structure \citep{Goldstein_2024}. The Gamma-ray Data Tool\rev{s} (GDT) packages extract data from the NASA High Energy Astrophysics Science Archive Research Center (HEASARC),\footnote{For \vxm\ the relevant data can be found at: \url{https://heasarc.gsfc.nasa.gov/FTP/fermi/data/gbm/triggers/2019/bn191117006/}.} enabling the time-binning of the data, extraction of energy-integrated flux, and time-integrated energy, for generating the lightcurve and energy spectra. \rev{For \rev{GRB\;191117A}, the prompt emission was detected by multiple GBM NaI scintillation detectors (with NaI 6 and NaI 7 being the only two significant events), which were therefore used in the subsequent analysis. W}e integrate the energy over $8\;{\rm keV}\,$--$\,900\;{\rm keV}$ to avoid the lower energy ranges where the GBM response is poorly calibrated, and to avoid the overflow channel in the higher energy ranges.

We also use the \texttt{gdt-fermi} package to fit a line to \rev{the time intervals before and after the \fermi\ signal}, in order to create a background subtracted lightcurve and spectra. While higher-order polynomials can be used with \texttt{gdt-fermi}, there becomes a risk of overfitting, so we utilize the simplest model for background subtraction.

\subsection{Swift} \label{sec:swift}

The \swift\ observatory \citep{Gehrels_2004} was primarily commissioned to detect and localize GRBs\footnote{The highest energy range that \swift\ can detect is hard X-rays, therefore, a significant portion are X-ray Bursts (XRBs). For the purposes of this paper define XRB to have a peak photon energy below $100\;\rm keV$.\label{foot:xrb}}, and within $\sim$90$\rm\,s$ slew the observatory for X-ray and ultraviolet/optical observations. \swift\ is unique in that all three of the telescopes are focused on detecting photons at higher energy levels than the optical wavelengths. It is one of the only space telescopes in operation that can detect both ultraviolet, with the use of the UltraViolet/Optical Telescope (\textit{UVOT}); and X-rays, with the X-ray Telescope (\textit{XRT}).

For this analysis, \vxm\ has data from both the \swift\textit{-UVOT} and \swift\textit{-XRT} telescopes. The \textit{Swift-XRT} data was taken several days after the \textit{Fermi-GRB} telescope data. The \swift\textit{-XRT} data, however, only provided upper bounds on the activity whilst the \fermi\ data \S\ref{sec:fermi} may provide X-rays at the time of first light.

The ultraviolet and bluer optical wavelength ranges help to further constrain SN luminosity around peak brightness. \swift\textit{-UVOT} has six filters with effective central wavelengths ranging from $2076\,\AA$ to $5412\,\AA$, covering the ultraviolet to visual bands. The \swift\textit{-UVOT} data were analyzed using the pipeline of the Swift Optical Ultraviolet Supernova Archive (SOUSA; \citealp{Brown_2014}), including the updated zeropoints of \citet{Breeveld_2011} and an aperture corrections based on the average UVOT PSF. These Vega magnitudes were converted to the AB system.



\subsection{Gaia}\label{sec:gaia}

Unlike \texttt{TESSreduce} for \tess, \atlas, \ps, and \ztf, \gaia\ does not perform difference-imaging in the traditional sense, but instead relies on precise astrometric measurements and source detection methods \citep{Gaia_2016}. \gaia\ contains multiple filters, but relies on the use of its broadband \gaia\textit{-G} filter for most observations. We convert the \gaia\textit{-G} band from the Vega system to the AB magnitude system.


\subsection{LCOGT}\label{sec:lcogt}


We also make use of data from the Las Cumbres Observatory Global Telescope Network \rev{as a part of the Global Supernova Project} (\lco\ \citep{Brown_2013} which has a series of 12 different $1\,{\rm m}$ photometric telescopes across five different continents\footnote{The photometry data for \vxm\ comes from McDonald Observatory in Texas, USA.} equipped with the \textit{Sinistro} cameras. The \lco\ images are processed using the \texttt{BANZAI} pipeline \citep{McCully_2018}.

\section{Lightcurve comparison}\label{sec:analysis}

\begin{figure*}
    \centering
    \includegraphics[width=1\linewidth]{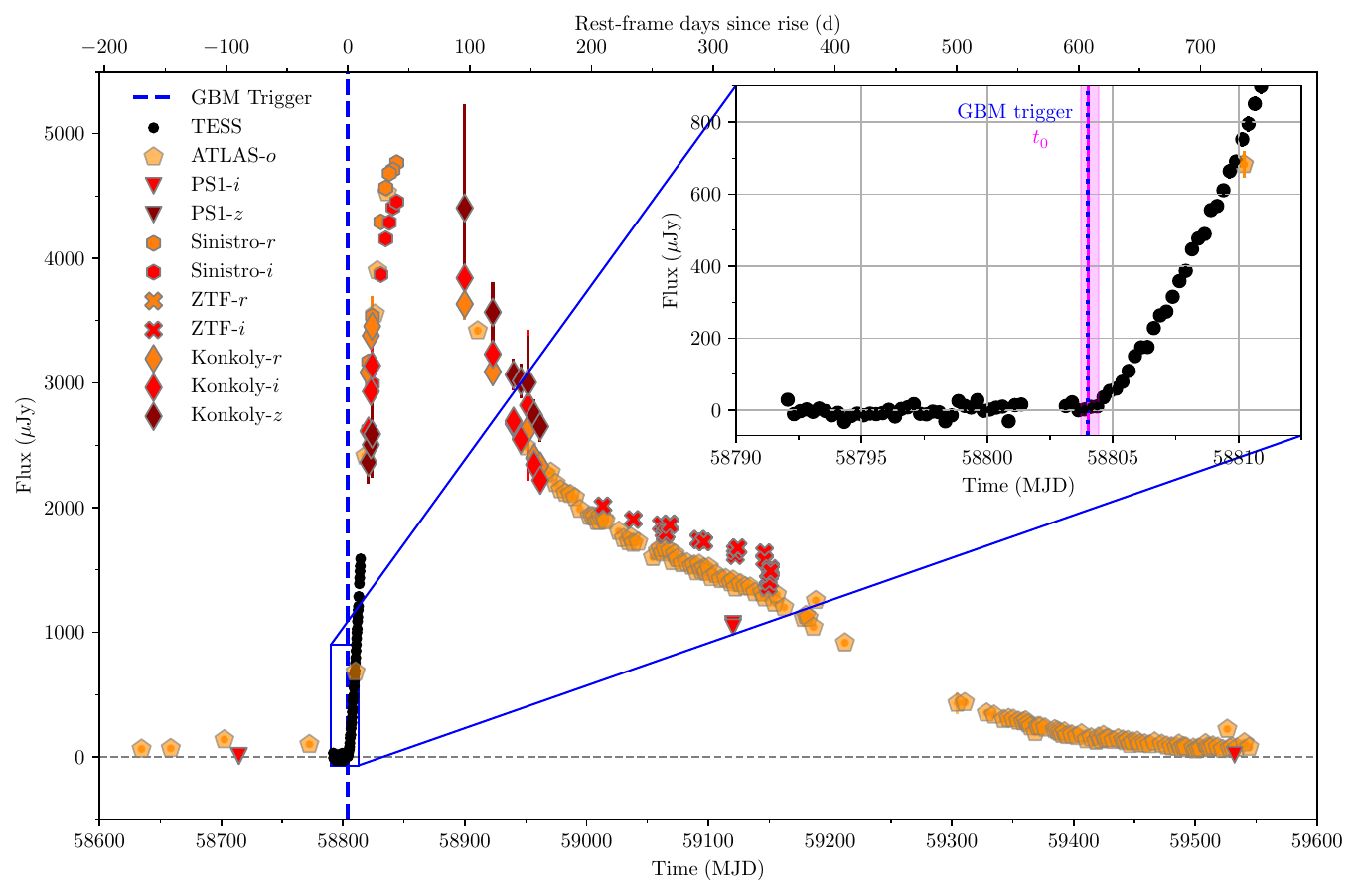}
    \caption{The `redder' filters are plotted in flux-density space to show the GBM trigger alongside the complete rise and decay. The \atlas\ data has been binned daily, and the \tess\ data is binned in groups of $6\,{\rm hr}$. We do not take into account band specific extinction. We note that most of the errorbars are not large enough to be visible. The vertical shaded region in the inset panel refers to the $1\sigma$ error around the \tess\ time of first light.}
    \label{fig:complete_rise}
\end{figure*}

\begin{figure*}
    \centering
    \includegraphics[width=1\linewidth]{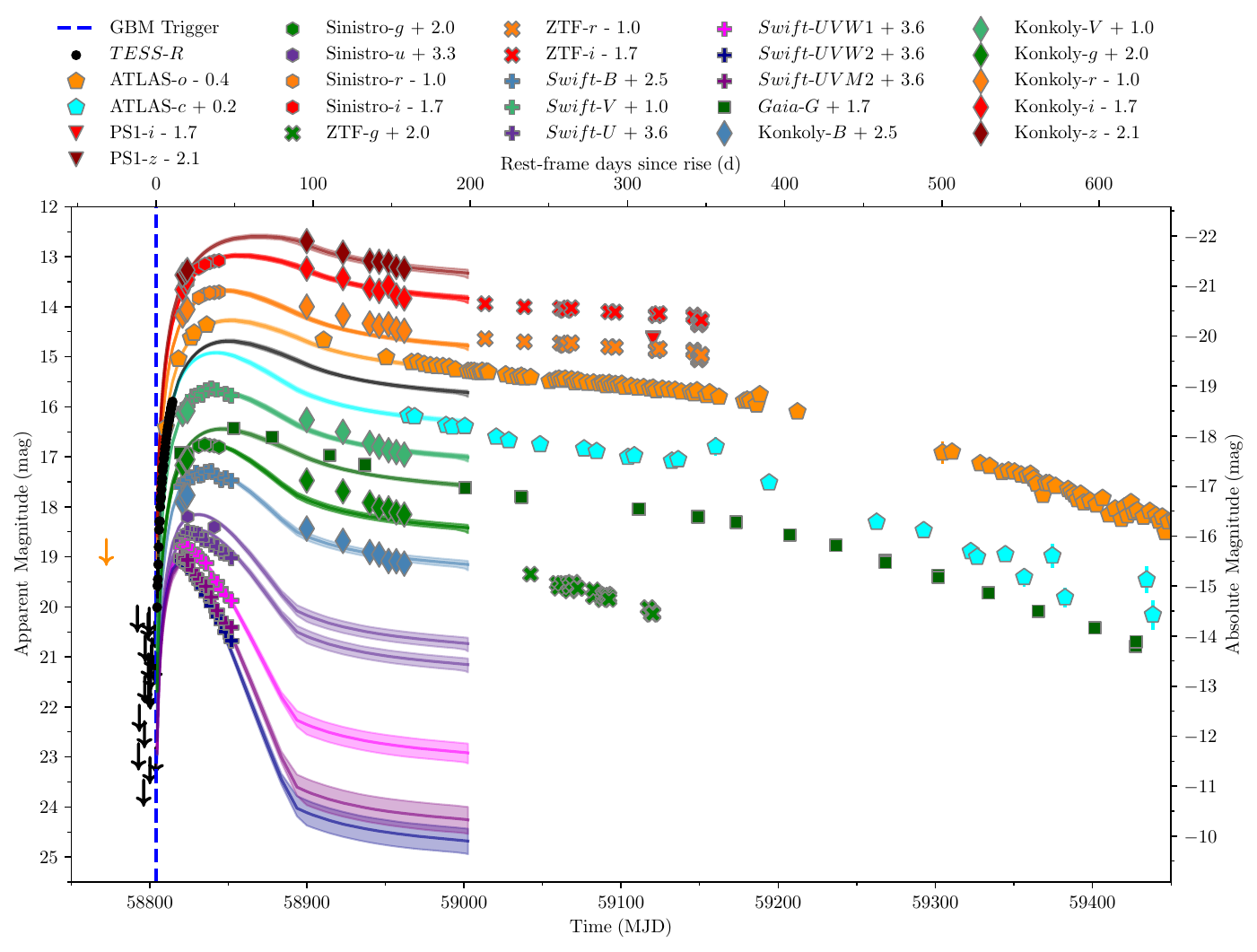}
    \caption{The complete lightcurve in all of the different bands plotted with different magnitude offsets for display purposes. The \atlas\ data has been binned daily, and the \tess\ data is binned in groups of $6\,{\rm hr}$. The plotted lines and the respective shaded regions represent the \mosfit\ lightcurve fitting median and the 5\% and 95\% confidence region for each of the respective bands fit within the first 200 days. For further discussion and details for the \mosfit\ fitting refer to \S\ref{sec:models}. For this data we have not corrected for extinction in the magnitudes. We note that most of the errorbars are not large enough to be visible.}
    \label{fig:magnitude}
\end{figure*}

We compare the light curve of \vxm, shown in \autoref{fig:complete_rise} and \autoref{fig:magnitude}, with those of several well-studied \iin\ \sne, as this can tell us a lot about the extent and density of the CSM and the central power source. To enable consistent comparison across different instruments and detection limits, we measure the time required for each event to fade by two magnitudes in a red-like band from peak. Under this definition, \vxm\ exhibits a relatively long decay time: it declines by two magnitudes in $\sim$375$\;\rm days$.

In contrast, many \iin\ \sne\ --such as SN\,2003ma \citep{Rest_2011}, SN\,2005gl \citep{Gal-Yam_2007, Gal-Yam_2009}, SN\,2005ip \citep{Smith_2009, Fox_2010, Katsuda_2014, Fox_2020, Shahbandeh_2024}, SN\,2006gy \citep{Smith_2007, Ofek_2007}, SN\,2006tf \citep{Smith_2008}, and SN\,2017hcc \citep{Moran_2023} -- fade by two magnitudes in $\lesssim 200$ days. Instead, \vxm\ more closely resembles the long-decaying events, which require  $\sim 200 \to 500\;\rm days$ to reach the same decline, such as SN\,2008am \citep{Chatzopoulos_2011}, SN\,2010jl \citep{Smith_2011, Ofek_2014_xray, Ofek_2019}, and SN\,2015da \citep{Smith_2024_15da}. This extended decay time likely reflects a more massive and radially extended CSM.


\rev{Many \iin\ \sne\ have been detected in X-rays, either through targeted observations or a coincidental pointing. In most cases, however, these detections are limited to relatively `soft' to `harder' X-ray energies, typically between $0.5\,\rm keV$ and $8\,\rm keV$. Primarily, this is a consequence of instrumental limitations with most X-ray observatories not being sensitive to energies above $\sim$8$\,\rm keV$. While the definition of ``hard'' X-rays is somewhat object- and context-dependent, emission at energies $\gtrsim$10 keV remains largely unexplored for the majority of \iin\ \sne.}

While SN\,2005ip, SN\,2006gy, SN\,2010jl, SN\,2017hcc all had X-rays observations spanning from days to years after the explosion trigger \citep{Katsuda_2014, Ofek_2014_xray, Smith_2007, Smith_2017_ip, Fox_2020, Chandra_2022}, \vxm\ could have one of the first X-rays detected from the trigger of the first light, making the possible detection rather special. The energy spectrum for \vxm\ between $-2$ and $9$ seconds is distinctive, peaking at \rev{$E_{\rm peak}^{}=78.6^{+23.2}_{-19.4}\;{\rm keV}$ but extending up to $\sim$120$\,\rm keV$, depending on the adopted spectral model and binning strategy}, higher than most X-ray detections for \sne.\footnote{It should be noted that the limit for most X-ray telescopes is in the soft X-ray regime, which limits the detectability for hard X-rays/more compact progenitors.}\,\footnote{During the lightcurve decay, the X-ray spectrum for SN\,2010jl did extend up to $\sim$50$\,\rm keV$, despite peaking at $\sim$1$\,\rm keV$, while the other SN events were limited to $8\,\rm keV$.} \rev{We adopted a Comptonized model from \texttt{gdt-fermi} as that gave the lowest $\chi^2/{\rm d.o.f.}$, however, the peak energy does not vary greatly between different photon flux models. If we simultaneously forward-model and constrain a potential higher-energy component using the non-detection from BGO 1 we result in a peak energy of $\sim$100$\,\rm keV$.}.


The absolute magnitude for \vxm\ gathered from \citet{Tsvetkov_2024}, which had good coverage of the peak, is $M_V^{}=-21.41\pm0.05\;{\rm mag}$ when corrected for extinction, but without $K$-corrections. The extinction correction, $A_V^{} = 1.29 \pm 0.32$, combines the contributions and errors from the Host Spectral Energy Distribution (described in \S\ref{sec:seds}); the insterstellar medium and local CSM environment along the line-of-sight, inferred by the relation between host hydrogen column density and extinction given by \citet{Guver_2009} using the \mosfit\ values; and Milky Way reddening, estimated from the dust map of \citet{Schlafly_2011}. With this correction, \vxm\ ranks among the most luminous subcategory of \iins\ alongside SN\,2003ma \citep{Rest_2011}, SN\,2006gy \citep{Smith_2007, Ofek_2007}, SN\,2006tf \citep{Smith_2008}, SN\,2008am \citep{Chatzopoulos_2011}, SN\,2010jl \citep{Smith_2011, Ofek_2014_xray, Ofek_2019}, SN\,2015da \citep{Smith_2024_15da}, and SN\,2017hcc \citep{Smith_2020_hcc, Chandra_2022, Moran_2023}. 

\begin{table}
    \centering
    \caption{The peak time and magnitude relative to the time since explosion from the \tess\textit{-R} band data. Minor band-specific offsets in the first light time are considered to be negligible. All magnitudes are not corrected for extinction.}
    \label{tab:peaks}
    \begin{tabular}{ccccc}
    \hline
        Filters\footnote{The prefix ``\textit{S-}" refers to the \swift\textit{-UVOT} telescope.} & Rise Time & App. Mag.\footnote{Not corrected for extinction.\label{tab1note}} & Abs. Mag.\footref{tab1note}\footnote{Despite not being in the Hubble flow, we adopt a fiducial spatially-flat $\Lambda$CDM cosmology with $H_0^{}=70\;{\rm km\,s^{-1}Mpc^{-1}}$ and $\Omega_{\rm M0}^{} = 0.3$.} & $\delta$Mag\\
         & days & AB Mag & AB Mag & \\
    \hline
        $g$ & $31.0^{+1.5}_{-2.0}$ & 14.748 & -19.836 & 0.005\\
        \textit{S-V} & $35.0^{+1.8}_{-1.7}$ & 14.64 & -19.94 & 0.06\\
        \textit{S-B} & $35.0^{+1.8}_{-1.7}$ & 14.78 & -19.80 & 0.05\\ 
        \textit{S-U} & $19.8^{+1.6}_{-0.7}$ & 14.91 & -19.67 & 0.04\\ \textit{S-UVW1} & $18.4^{+0.7}_{-0.5}$ & 15.17 & -19.41 & 0.04\\
    \hline
    \hline
    \end{tabular}
\end{table}

\section{Shock breakout \rev{Analysis}}\label{sec:shocks}

\subsection{Optical precursor and Shock Checks}

\iin\ \sne\ often exhibit ``precursor'' events, associated with extreme mass-loss \citep{Smith_2014, Ofek_2014_precursor}. These outbursts are more prevalent in the period of months prior to the time of explosion, but can take place over years or even centuries \citep{Smith_2010, Mauerhan_2013, Ofek_2013_outburst, Ofek_2014_precursor, Bilinski_2015, EliasRosa_2018}. 

These mass-ejection events, combined with the assumption that the early lightcurve rise is powered predominantly by shock interaction, suggests potential correlations between the precursor's integrated luminosity and the SN's peak luminosity and rise time \citep{Ofek_2014_precursor}. For \vxm\ we attempt to find any evidence for a precursor event, which may persist for several months.

We test the \atlas\ and \tess\ data up to the time of the explosion to search for any precursor-like events. For the \atlas\ data we take 150 days prior to the time of first light, while for \tess\ we only have the month prior.

To find the significance for any precursor or shock breakout events across a range of timescales, we utilize \texttt{ATClean} \citep{Rest_2025} to minimize artifacts and detect low-level transient emission above zero flux in the \atlas\ lightcurve. We can apply a weighted Gaussian rolling sum with different kernel sizes $\sigma_\text{kernel}^{}$ = 0.1, 0.2, 0.4, 0.6, 0.8, 1, 2, 3, and 5 days, matched to---and thus amplifying---precursor timescales of interest. Our key figure of merit, $\Sigma_\text{FOM}^{}$, is thus defined as the flux-to-uncertainty ratio convolved with the
weighted rolling Gaussian sum. This allows enhancement of faint, extended emission episodes and makes them distinguishable from noise.

To determine whether a detected signal is real, we establish dynamic detection limits $\Sigma_\text{FOM,limit}^{}$ by analyzing control light curves known to contain no astrophysical flux. For each $\sigma_\text{kernel}^{}$, we adjust the detection limit to allow no more than two control light curves with false positives. If any $\Sigma_\text{FOM}^{}$ exceeded this limit in the SN light curve, the signal is flagged as a candidate outburst. This conservative detection limit ensures high purity while maintaining good sensitivity to faint outbursts.

\begin{figure}
    \centering
    \includegraphics[width=1\linewidth]{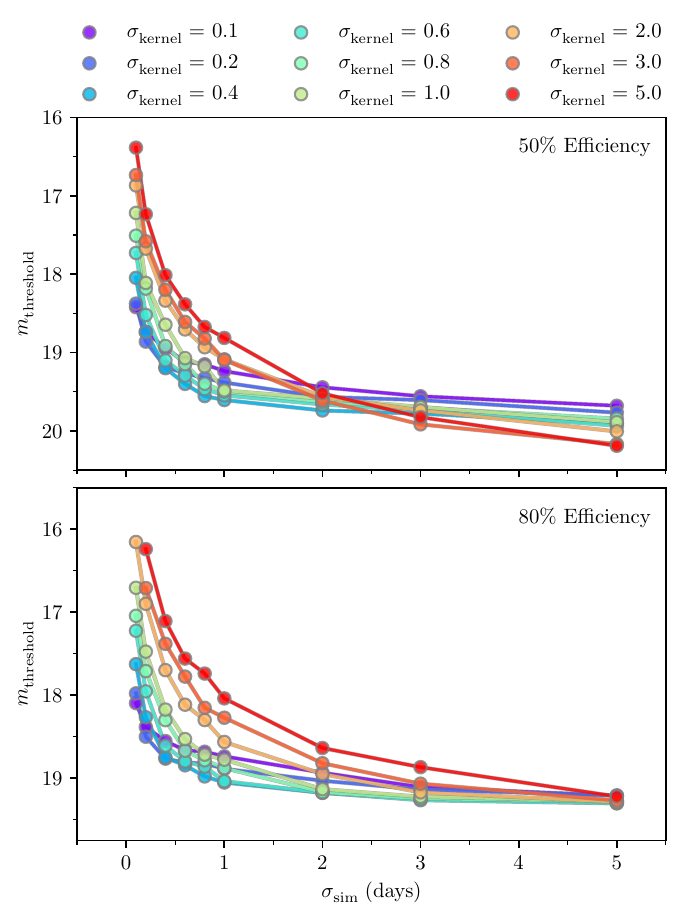}
    \caption{The detection limits for the precursor search in \tess\ with injected/simulated gaussian signals, $\sigma_{\rm sim}^{}$ (days), tested against Gaussian kernels of varying widths (days).}
    \label{fig:precursor}
\end{figure}

We additionally evaluate the detection algorithm's sensitivity by injecting many synthetic precursor or shock breakout events--both modeled as Gaussian bumps--into the control light curves and counting how many were successfully detected. By varying event timescale (i.e., Gaussian $\sigma_\text{sim}^{}$) and peak apparent magnitude ($m_\text{peak}^{}$), we can quantify detection efficiency as a function of those parameters. We repeat the process across all $\sigma_\text{kernel}^{}$. Lastly, we use the resulting efficiency curves to determine the 50\% and 80\% magnitude detection thresholds for each precursor timescale $\sigma_\text{sim}^{}$. The detection efficiency is displayed in \autoref{tab:precursor} and visualized in \autoref{fig:precursor}.

We also exclude several MJD ranges from our analysis if very bright injected simulations do not reach 100\% efficiency. First, we exclude gaps between observation seasons, as the lack of measurements makes both real and simulated precursor detection nearly impossible. Second, we add a buffer of 1 excluded day to our first observation season after observing an edge effect with significantly more nondetections very close to the upcoming season gap. 

Ultimately, no statistically significant precursor emission or shock breakouts were detected prior to the $\sim$5 magnitude rise to peak in the \vxm\ light curve across the tested timescales in either \atlas\ and \tess, allowing us to place stringent constraints on outburst brightness and frequency for the progenitor.\\
\\
\subsection{Localization Probability}\label{sec:loc}

\begin{figure}
    \centering
    \includegraphics[width=1\linewidth]{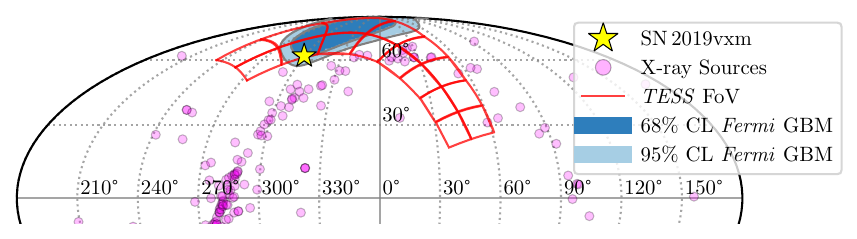}
    \caption{A Northern hemisphere RA and Dec Mollweide projected skymap showing the position of \vxm, the asymmetric 68\% and 95\% confidence level (CL) regions from the probability distribution map, known X-ray binary positions, and an overlay of the \tess\ pointing for Sector 18. The \tess\ field covers both $96.7\%$
    and $81.9\%$ of the 68\% and 95\% CL error regions respectively.}
    \label{fig:sky_map}
\end{figure}

\begin{table*}
    \caption{A numerical display of \autoref{fig:precursor} to display the magnitude thresholds for the optimal kernel size and the rate that simulated gaussian signals are recovered at these thresholds.}
    \label{tab:precursor}
    \centering
    \begin{tabular}{ccccc}
        \hline 
        \hline 
        $\sigma_{\rm sim}^{}$ & Best App. Mag. Threshold $50.0\%$ & Best $\sigma_{\rm kernel}^{}\;{\rm at}\;50.0\%$ & Best App. Mag. Threshold $80.0\%$ & $\sigma_{\rm kernel}^{}\;{\rm at}\;80.0\%$ \\
        \hline 
        0.10 & 18.42 & 0.10 & 18.09 & 0.10 \\
        0.20 & 18.86 & 0.20 & 18.50 & 0.20 \\
        0.40 & 19.20 & 0.40 & 18.76 & 0.20 \\
        0.60 & 19.40 & 0.40 & 18.84 & 0.40 \\
        0.80 & 19.55 & 0.40 & 18.98 & 0.40 \\
        1.00 & 19.61 & 0.40 & 19.05 & 0.40 \\
        2.00 & 19.74 & 0.40 & 19.18 & 0.40 \\
        3.00 & 19.92 & 3.00 & 19.26 & 0.60 \\
        5.00 & 20.19 & 5.00 & 19.30 & 0.60 \\
        \hline 
    \end{tabular}
\end{table*}

To determine the probability for whether \vxm\ and the \fermi\ trigger alert \rev{GRB\;191117A} are the same event or are merely a coincidence, we perform the following tests:
\begin{enumerate}
    \item Find if there are any other optical triggers within a time frame of $\sim$3$\,{\rm days}$ using the \tess\ field.
    \item Determine if there are any known X-ray sources, such as X-ray binaries, within $2\sigma$ of the GRB trigger.
    \item Find the spatial and temporal probability that the two are correlated.
\end{enumerate}

We run the \texttt{TESSELLATE} pipeline \citep{Roxburgh_2025} on the sector 18 field of \tess\ to try and find any transient-like event within a $2\sigma$ radius of the reported location of \rev{GRB\;191117A}. The wide FoV of \tess\ covers $96.7\%$ and $81.9\%$ of the 1 and 2$\sigma$ error regions respectively for \rev{GRB\;191117A} (refer to \autoref{fig:sky_map} for a visual representation). Injection and recovery tests for the \texttt{TESSELLATE} pipeline show $\sim$50$\%$ successful recovery rates up to $m_{\rm TESS}^{}\sim 17$, and no event other than \vxm\ was detected. This could mean that either a hypothetical coincident event (i) was not bright enough in the \tess\textit{-R} band, (ii) the optical component which was a prompt emission/evolved quickly in time, e.g., lasted less than 30 minutes, or (iii) was not covered within the \tess\ field, i.e., outside the observed region. Nevertheless, the fact that \vxm\ is the only detected transient within the error region makes it a strong candidate for association with the XRB (see Footnote~\ref{foot:xrb} for the definition of XRB in this context).

\subsubsection{Spatial Coincidence}

In addition to the optical analysis, we compared the spatial position of \rev{GRB\;191117A} with entries in the \texttt{XRBcats} X-ray binary catalog\footnote{\url{http://astro.uni-tuebingen.de/~xrbcat/}} \citep{Avakyan_2023, Neumann_2023} to exclude an association with any known repeating X-ray source. As \texttt{XRBcats} lists no known X-ray binary within the $2\sigma$ error region of \rev{GRB\;191117A}, this supports the identification of \vxm\ as a likely source.

To estimate the spatial probability of two events being co-located, we utilize the \texttt{gdt-fermi} package \citep{Goldstein_2023} to load the probability localization \texttt{HEALPix} map \citep{Gorski_2005}. The map yields a lower-bound probability for the spatial discrepancy of $P_{\rm diff,\,loc}(\vec{x}) = 2.5\%$. Using the \fermi\ probability distribution, we evaluate the likelihood of a chance spatial association under the null hypothesis that the sources are unrelated. We load all \fermi\ GRB skymaps with \texttt{healpy} and, for each GRB, generate 10,000 random sky positions. We then sample the interpolated probability at each position to construct a null (background) distribution. This distribution provides the reference against which we compare the observed sky positions and associated probabilities of \rev{GRB\;191117A} and \vxm. These simulations yield the probability the two events are the same based on their spatial location alone as $P_{\rm diff,\,loc}(\vec{x}) = 0.61\%$.


\subsubsection{Temporal Coincidence}

When estimating the probability of temporal coincidence, $P_{\rm diff}(t)$, care must be taken in defining the conditioning framework. We considered several formulations for computing the chance coincidence between a SN and an X-ray burst (XRB):
\begin{enumerate}
    \item For a fixed SN, what is the probability of a XRB happening within a given timeframe or vice versa,
    \item A symmetric probability formulation upon accounting for a fixed event Y given at least one event Z for both \sne\ and XRB,
    \item The probability of a chance-coincidence that these two events occur within a given timeframe under null-hypothesis simulations,
    \item What is the probability that a SN and a XRB happen within a certain specified fixed window timeframe.
\end{enumerate}
Each approach is valid under specific assumptions, but we focus on those most consistent with our physical scenario and statistical requirements.

\medskip

\textbf{Option 1 --- Asymmetric Conditioning:} 
This approach assumes one fixed event (e.g., a known SN), and evaluates the likelihood of a second event (e.g., XRB) within a window $\delta t$, inherently being asymmetric in formulation. The choice of which rate to use, and what assumption to enforce is highly non-trivial. As the observed \sne\ rate is greater than the observed XRB rate, an argument can be made that the Poisson probability \rev{$P(\geq1\,S|\geq1\,X,\,\delta t)_{\rm Poiss.}^{}=1-\exp(-\delta t\,R_{S}^{})$} should be used, where $R_{S}^{}$ is the SN rate and $\delta t $ is the time-range of the coincidence window. As the framing of this paper is with respect to the SN itself, conversely, the inverse assumption could be made, \rev{$P(\geq1\,X|\geq1\,S)_{\rm Poiss.}^{}=1-\exp(-\delta t\,R_{X}^{})$}, however, both formulations are inherently asymmetric and likely misrepresent the true coincidence.

\medskip

\textbf{Option 2 --- Rate Symmetric Conditioning:}
To address the asymmetry, we compute a generalized coincidence probability weighted by the respective event rates:
\begin{equation}\label{eq:poisson}
    P_{\rm diff,\,theory}^{}(t) = \frac{R_{\rm S}^{} P(X|S,\,\delta t) + R_{\rm X}^{}P(S|X,\,\delta t)}{R_{\rm S}^{} + R_{\rm X}^{}} \; ,
\end{equation}
where $R_{\rm Y}^{}$ is the rate of either XRBs or SNe within a given timeframe, and \rev{$P(Y|Z,\,\delta t)= P(\geq1\,Y|\geq1\,Z,\,\delta t)_{\rm Poiss.}^{}$} is the Poisson probability of an event $Y$ happening given at least one event $Z$ happens within the uncertainty $\delta t$.

\medskip

\textbf{Option 3 --- Chance-coincidence Simulations:}
The third methodology is most effectively implemented using a brute-force simulation to test the null hypothesis, similar to the approach of \citet{Moroianu_2023}. We generate thousands of synthetic samples of SN and XRB events spanning a 14-year period to estimate the frequency with which the two event types occur within a time interval $\delta t$. The empirical probability is defined as
\begin{equation}
    P_{\rm diff}(t) = \frac{ N_{\geq 1,\,\rm S}^{\rm X} + N_{\geq 1,\,\rm X}^{\rm S}}{N_{\rm S} + N_{\rm X}} \; ,
\end{equation}
where $N_{\geq 1,\,Z}^{\rm Y}$ is the number of Y-type events that temporally overlap with at least one Z-type event within $\delta t$, and $N_{\rm Y}$ is the total number of Y-type events, drawn from a Poisson distribution. Here, $\rm Y$ and $\rm Z$ represent either SN or XRB events within the simulated timeframe, with rates $R_S$ and $R_X$. We adopt this simulation-based estimate as our fiducial value, as it minimizes assumptions and fully captures the time-domain structure of the data. Notably, the simulated result is consistent with the Rate Symmetric Conditioning probability, confirming agreement between the two methods.

\medskip

\textbf{Option 4 --- Fixed-Window Coincidence:}
This formulation estimates the probability that the two separate events occur within a predefined time window, as opposed to integrating over all possible time windows $\delta t$. This is done computationally by simulating XRB and \sne\ and counting the overlaps in time. It generally underestimates the true coincidence probability and is considered the least likely choice. \\

\begin{table}
    \centering
    \caption{The probability for the temporal association of the XRB and \vxm.}
    \label{tab:prob}
    \begin{tabular}{ccc}
    \hline
        Method & Probability of Coincidence & $\sigma$ CL \\
    \hline
        Option 1, XRB rate & 11.53\% & 1.6 \\
        Option 1, SN rate & 30.66\% & 1.0\\
        Option 2 & 15.35\% & 1.4 \\
        Option 3 & 15.32\% & 1.4 \\
        Option 4 & 7.36\% &  1.8\\
    \hline
    \hline
    \end{tabular}
\end{table}

For the SN rate, we adopt the empirical observed rate for core-collapse SN (using the observed fraction of 0.21 of all SN from \citep{Li_2011, Arcavi_2017}) averaged over the last three years reported in the TNS (see Footnote~\ref{foot:tns}). Close distance volumetric rates can lead to significant overestimation, as SN brightness and detection efficiency can fall off dramatically beyond $z_{\rm hel.}^{}\sim0.1$, making redshift cuts both necessary and difficult to implement—particularly given the wide range of possible absolute magnitudes.

For the XRB rate we initially make the conservative over-estimate that 50\% of all GRBs are actually XRBs, and estimate the rate using 14 years of GRB data between 2011 and 2024.\footnote{\url{https://user-web.icecube.wisc.edu/~grbweb_public/}} We later marginalize the XRB fraction to minimize the impact of this assumption.

Using $P_{\rm diff.}^{}(\vec{x}) = 0.61\%$ and $P_{\rm diff.}^{}(t) = 15.3\%$ (from option 3), the probability that the two events are not independent is $P_{\rm same}^{}(\vec{x}, t) = 1- P_{\rm diff.}^{}(\vec{x}) \cdot P_{\rm diff.}^{}{(t)} = 99.91\%$, corresponding to a $3.3\sigma$ confidence level.

We note that many of the values used potentially overestimate the coincidence-chance, resulting in a conservative estimate of the probability that the two events are associated.

Instead of using a fixed XRB fraction of total GRBs, we marginalize over the parameter for all options shown in \autoref{tab:prob}, giving us $P_{\rm same}^{}(\vec{x}, t) = 99.91\%$, also corresponding to a $3.3\sigma$ confidence level, with $95\%$ and $5\%$ XRB fraction constraints being between $3.2\sigma$ and $3.8\sigma$. A more robust temporal coincidence probability calculation would require information about the physical system and the expected time offset between the SN and the XRB to test the null hypothesis, although this would require a great knowledge of the asymmetric 3D CSM structure and the expected time delay. However, despite estimates from \autoref{sec:models}, this is beyond the scope of the paper.

While the probability that the two events are physically associated is high, with only a $0.09\%$ chance of coincidence, the uncertainties—particularly the lack of radial distance information, 3D CSM structure, and reliance on the \tess\ detection limit—prevent a definitive claim. Therefore, we conclude that the two events are most likely associated.\\

\section{Modeling}\label{sec:models}

Modeling the lightcurve for a Type IIn SNe can reveal important aspects about what the possible progenitor may be, as well as the environment, and other possible physical factors at play. The luminosity and shape profile can tell us about the interaction with the CSM and the mass-loss history for the progenitor/system \citep{Arcavi_2017, Smith_2014, Smith_2017_book}. Both the luminosity and the duration of the SN during the photospheric phase rely strongly on the diffusion properties of the radiation through the CSM, therefore, they can be used to infer the structural properties of the larger CSM \citep{Villar_2017}. Additionally, further color information can potentially constrain the optical density of the CSM and provide information about the progenitor.

\subsection{Model used and Assumptions}\label{sec:model_assume}

Modeling Type IIn \sne\ is challenging due to the degeneracies associated with the CSM (refer to \S\ref{sec:model_discuss} for more details). To address this, we attempt to model \vxm\ using CSM interaction models \citep{Chatzopoulos_2012, Chatzopoulos_2013, Villar_2017, Jiang_2020}, which were built upon the earlier pioneering work of \citet{Arnett_1980, Arnett_1982}. These models simplify the complex dynamics using the following assumptions:
\begin{enumerate}
    \item The expansion of the ejecta is homologous.\footnote{The homologous solution assumption has been allowed to relax in the work of \citet{Chatzopoulos_2012, Chatzopoulos_2013}.}
    \item The power/heat source is centrally located.
    \item The radiation pressure dominates over electron or gas pressure in the Equation of State (EoS).
    \item There is negligible energy loss due to neutrinos.
    \item The mathematical differential equation assumes separability of spatial and temporal behavior.
    \item The system is spherically symmetric.
    \item The main energy engine is the conversion of kinetic energy to radiation as the shock impacts the dense CSM,  while contributions from $^{56}$Ni and $^{56}$Co radioactive decay are negligible.
\end{enumerate}
Due to the likely asymmetries in the ejecta as discussed in \S\ref{sec:asym}, we adopt the values as illustrative of the progenitor properties to contextualize this event relative to other events,  rather than an accurate determination of them.


\subsection{Bayesian Modeling}\label{sec:bayes_model}


To implement the CSM interaction models described in \S\ref{sec:model_assume}, we use the Bayesian inference \texttt{python} package \mosfit\footnote{\url{https://github.com/guillochon/MOSFiT}} \citep{Guillochon_2018}. Specifically, we use the \texttt{dynesty} nested sampling package \citep{Higson_2019, Speagle_2020} for the \mosfit\ models to explore the posterior distribution.

In Bayesian statistics, the \textit{posterior} probability is computed as the ratio of the likelihood and prior to the evidence, given by:
\begin{equation}
    \mathcal{P}(\Theta|d,\,M) = \frac{\mathcal{L}(d | \Theta,\,\mathcal{M})\,\mathcal{P}(\Theta | \mathcal{M})}{\mathcal{P}(d | \mathcal{M})} \; ,
\end{equation}
where $\Theta$ are the model parameters, $d$ the data, $\mathcal{M}$ the model assumed. The likelihood\footnote{Here, we use a Gaussian likelihood with quadrature noise to account for extended error regions.} can be written as the $\mathcal{P}(d | \Theta, \mathcal{M})$ term, and $\mathcal{P}(\Theta | \mathcal{M})$ represents the prior $\pi$, and $\mathcal{P}(d | \mathcal{M})$ is the evidence $\mathcal{Z}$ of model $\mathcal{M}$.

Since Bayesian sampling relies on prior information, the choice of prior distribution is crucial to obtaining a valid posterior. To be consistent with other notable \iin\ fitting analyses we adopt the same priors as \citet{Ransome_2025}, and fix the redshift as $z_{\rm hel.}^{}=0.019$.

To ensure accurate modeling of the posterior distribution and to take in effect the high number of expected degeneracies in the multidimensional parameter space, we choose a high number of `live points', $n_{\rm live}^{} = 1000$, with a tolerance of $d \ln \mathcal{Z} = 10^{-1}$ to match \citet{Ransome_2025}. We show the posterior parameter distribution in \autoref{fig:mosfit_corner}, while the lightcurve model can be seen in \autoref{fig:magnitude}.

\subsection{Host Spectral Energy Distribution}\label{sec:seds}

\begin{table*}
    \centering
    \caption{Comparing the galaxy SED of \host\ with other notable SLSN-\iin\ using the data from \citet{Schulze_2018}.}
    \label{tab:hosts}
    \begin{tabular}{ccccccc}
    \hline
        SN Name & Redshift & Host Metallicity & Host Stellar Mass & Host Age & Star Formation Rate & Specific SFR\\
        &  & $\log_{10}{\left(\frac{Z_*^{}}{Z_{\odot}^{}}\right)}$ & $\log_{10}{\left(\frac{\rm M}{\rm M_{\odot}^{}}\right)}$ & $\log_{10}{\rm yr}$ & $\log_{10}{\rm SFR}$ & $\log_{10}{\rm sSFR}$\\
    \hline
        \vxm\ & 0.019 & $-0.13^{+0.22}_{-0.26}$ & $7.72^{+0.37}_{-0.55}$ & $9.75^{+0.15}_{-1.10}$ & $-0.89^{+0.27}_{-0.33}$ & $-8.63^{+0.57}_{-0.52}$\\
        SN\,2003ma$\,$\footnote{\citet{Rest_2011}} & 0.289 & -- & $8.91^{+0.13}_{-0.12}$ & $7.63^{+0.24}_{-0.29}$ & $1.34^{+0.16}_{-0.13}$ & $-7.55^{+0.21}_{-0.24}$\\
        SN\,2006gy$\,$\footnote{\citet{Smith_2007, Ofek_2007}} & 0.019 & -- & $11.70^{+0.06}_{-0.21}$ & $9.91^{+0.10}_{-3.18}$ & $-1.12^{+0.08}_{-0.08}$ & $-6.78^{+0.12}_{-7.17}$\\
        SN\,2006tf$\,$\footnote{\citet{Smith_2008}} & 0.074 & -- & $7.54^{+0.47}_{-0.20}$ & $8.86^{+0.40}_{-0.35}$ & $-1.25^{+0.48}_{-0.37}$ & $-8.88^{+0.35}_{-0.29}$\\
        SN\,2008am$\,$\footnote{\citet{Chatzopoulos_2011}} & 0.233 & -- & $9.28^{+0.13}_{-0.15}$ & $8.57^{+0.33}_{-0.24}$ & $0.74^{+0.23}_{-0.24}$ & $-8.50^{+0.20}_{-0.35}$\\
        SN\,2008fz$\,$\footnote{\citet{Drake_2010}} & 0.133 & -- & $6.55^{+0.25}_{-0.28}$ & $8.64^{+0.71}_{-0.67}$ & $-2.08^{+0.47}_{-0.48}$ & $-8.64^{+0.71}_{-0.67}$\\
        SN\,2009nm$\,$\footnote{\citet{Drake_2009, Christensen_2009}} & 0.210 & -- & $8.65^{+0.33}_{-0.34}$ & $8.95^{+0.62}_{-0.52}$ & $-0.60^{+0.65}_{-0.62}$ & $-9.20^{+0.79}_{-0.83}$ \\
    \hline
    \hline
    \end{tabular}
\end{table*}



We model the host-galaxy Spectral Energy Distribution (SED) using photometry from the Sloan Digital Sky Survey (SDSS), AllWISE, and Pan-STARRS1.\footnote{Additional infrared data would further constrain the SED and help mitigate parameter degeneracies.} The host photometry and stellar population properties are determined through \texttt{FrankenBlast} \citep{Nugent_2025}, a rapid tool used for transient host association, host photometry, and SED modeling, and will be compared to the spectra host-galaxy analysis in \citet{Smith_prep}.

We classify the host as a dwarf galaxy with a total stellar mass of $M_* = 5.2^{+7.1}_{-3.8} \times 10^{7}\,\mathrm{M}_{\odot}$ and a stellar metallicity of $\log_{10}(Z_*^{}/Z_\odot^{}) = -0.13^{+0.22}_{-0.26}$. This places the galaxy below both the Small and Large Magellanic Clouds in mass, but with a metallicity comparable to that of the Large Magellanic Cloud \citep{Hocde_2023}. The galaxy shows little evidence of AGN activity and exhibits an elevated star formation rate over the past $\sim$30$\;\rm Myr$, increasing over an order of magnitude from $\sim$10$^{-2}$ to $\sim$10$^{-1}\,\mathrm{M}_{\odot}^{}\,\mathrm{yr}^{-1}$. 

\begin{figure}
    \centering
    \includegraphics[width=1\linewidth]{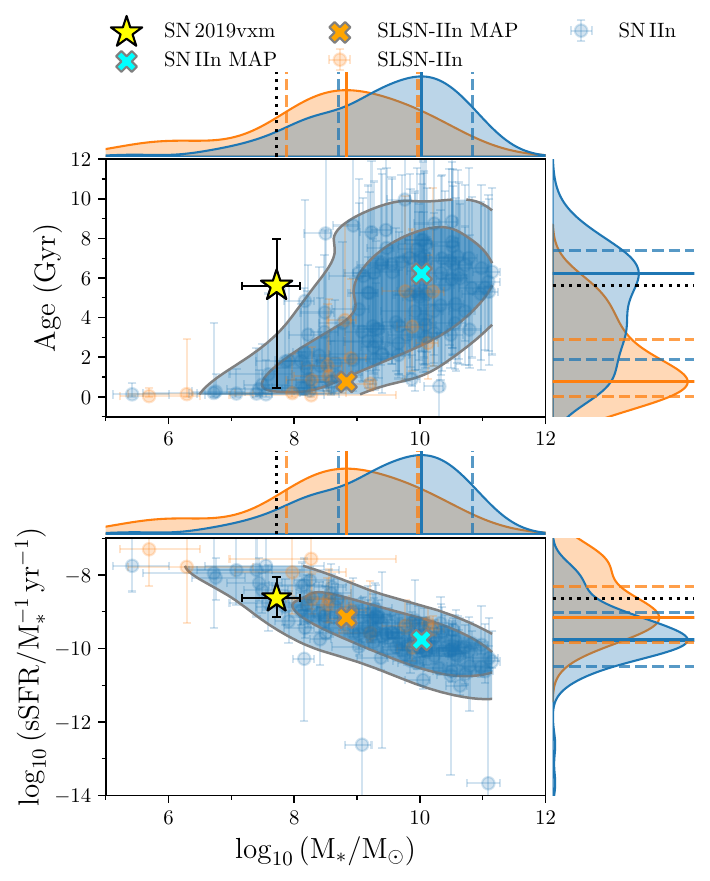}
    \caption{Host SED comparisons between the host of \vxm\ and those within the \citet{Schulze_2021} sample. \vxm\ is discrepant with the general \iin\ distribution by 1.8$\sigma$ for host age and stellar mass, and by 1.5$\sigma$ for host specific SFR and stellar mass. We have marked the Maximum A Posteriori (MAP) values of the posterior on the plot for comparison.}
    \label{fig:host}
\end{figure}

Two environmental trends have been reported for SLSN-IIn host galaxies: (i) they are typically dwarf systems ($M_*^{} \lesssim 10^{10}\,\rm{M}_{\odot}^{}$), with massive hosts such as that of SN\,2006gy being rare \citep{Schulze_2018}; and (ii) they display redder $R - K_s$ colors, indicative of older stellar populations and consistent with star-forming galaxies \citep{Schulze_2018}. The stellar mass of \vxm\ is consistent with the dwarf-galaxy regime; however, its $R - K_s$ color cannot be determined due to the absence of NIR observations. We show a comparison of different host properties for SLSN-IIn in \autoref{tab:hosts} and \autoref{fig:host} based on the data from \citet{Schulze_2018} and \citet{Schulze_2021} respectively. No robust conclusions can yet be drawn, given the small sample size and the diversity of host environments. While weak correlations have been found relating the overall luminosity to both low metallicity and a younger stellar age population \citep{Moriya_2023}, it is not necessarily apparent how strongly these impact the total luminosity of the SN. Nonetheless, \citet{Moriya_2023} find that higher mass-loss rates and wind velocities tend to occur in more metal-poor environments, which seems to be consistent with \vxm. This is, of course, the opposite of what is expected from radiation-driven winds, which should be harder to drive, and have lower $M_\odot^{}$ at low metallicity. In this case it is likely because the CSM is not from radiation driven winds. It is important to note that metallicity measurements using only photometry are difficult, with stronger constraints coming from spectra. The effect of metallicity explored more in  \citet{Nugent_2025} and potentially expanded in \citet{Nugent_prep}.

\section{\rev{Discussion}}\label{sec:discussion}

\subsection{Shock Breakout Discussion}\label{sec:sbo_discussion}

For the stellar surface, the higher energy/stronger emission of X-rays over short timescales (on the order of seconds) are often associated with relativistic surface shock breakouts \citep{Chevalier_2011, Chevalier_2012, Waxman_2017}, while, for slower breakouts, the X-rays are suspected to become dominant for longer at later times. Typically the shorter timescales and mid- to high energy X-rays correlate quite strongly with a smaller initial radius of the progenitor \citep{Waxman_2017}. Indeed, despite the high energy emitted by the explosion of a RSG, significant portions of the radiation are expected to be absorbed by the interstellar medium (ISM), with low--intermediate UV energies produced \citep{Waxman_2017}. Smaller-radius progenitors, such as BSGs and Wolf-Rayet stars, however, are expected to have shock energies in the medium--high X-ray range \citep{Waxman_2017}. Both soft and hard X-rays have already been observed for \iin\ \sne, such as SN\,2010jl \citep{Ofek_2014_xray, Chandra_2015}; however, these are typically from continued ongoing CSM interaction.

Not only can a shock be formed at the stellar surface, a more luminous shock breakout can be formed more commonly within the dense CSM around a \iin\ \sne\ progenitor \citep{Smith_McCray_2007}. If the Thomson optical depth, $\tau$, of the CSM is larger than $c/v_{\rm sbo}^{}$, where $v_{\rm sbo}^{}$ is the shock breakout velocity and $c$ is the speed of light, the shock will form within the CSM as opposed to the stellar surface \citep{Ofek_2014_interaction}. Due to the optically thick CSM, and assuming an approximately spherical structure, it is unlikely we will directly observe a stellar surface breakout. In fact, this circumstellar envelope can still significantly change the observable features including spectroscopic line widths, as well as the photon energy and duration for any CSM shock breakout event. 

In addition to the breakout itself, another mechanism for UV and X-ray emission in \iin\ \sne\ arises from ongoing interaction between the ejecta and the dense CSM. Radiatively-driven, optically thick winds that formed the CSM can give rise to a collisionless shock once the radiation-mediated shock has passed \citep{Ofek_2014_interaction, Moriya_2023}. These shocks can produce hard X-rays via inverse Compton scattering and other non-thermal processes. Unlike the shock breakout, which reflects the progenitor's structure, these are later-time emissions which are more sensitive to the CSM density, geometry, and shock physics. Detectors such as \fermi\ and \swift\ among others could possibly detect the X-rays from potential shock breakout events.

\subsection{\fermi\ GBM Discussion}\label{sec:asym}

The \fermi\ light curve precedes the magnitude-limited modelled \tess\ explosion by $0.61\,\mathrm{hrs}$ and exhibits a distinct spectral peak at \rev{$E_{\rm peak}^{}=78.6^{+23.2}_{-19.4}\;{\rm keV}$ for a simplistic Comptonized photon model. However, depending on the binning and reduction constraints the peak photon energy may be up to $120\;\rm keV$.}. \rev{The resulting} X-ray flash \rev{follows} the general population in the $E_{\rm peak}$–$E_{\rm iso}$ (Amati) relation \citep{Amati_2006, Heussaff_2013}, indicating that its spectral and energetic properties \rev{are consistent} with typical X-ray flashes and GRB jets. Such an energy range is consistent with a compact progenitor. Intermediate-mass blue supergiants (BSGs) are expected to produce photons primarily in the $1$---$10\,\mathrm{keV}$ range, while Wolf–Rayet stars typically emit in the $2$---$50\,\mathrm{keV}$ range, sometimes extending to $\sim$100$\,\mathrm{keV}$ \citep{Waxman_2017}.

However, the photon energy of the X-ray flash in the presence of CSM becomes more complex as it must encompass the properties of the shock, and either the density/geometry of the stellar surface or the CSM. The average photon energy produced within the shocked region during a breakout is unlikely to follow a blackbody distribution, and instead can be much larger than the temperature would typically dictate assuming thermal equilibrium \citep{Balberg_2011}. The optical depth from a steady wind alone will alter the observed dynamics and properties of the shock breakout \citep{Balberg_2011}.

For a standard shock breakout from a barren spherically-symmetric stellar surface, the emission timescale is expected to be short, $90\%$ of the integrated counts (T90) for the \vxm\ \fermi\ lightcurve are observed within $7.424\,\mathrm{s}$. Such brief durations generally imply a smaller progenitor radius \citep{Drout_2014}. However, the presence of a CSM can significantly alter the dynamics.

In a spherically symmetric, homogeneous CSM, the effective shock velocity is reduced compared to a direct surface breakout, following the relation $R_{\mathrm{csm}} \sim v_{\mathrm{sbo}}^{} t_{\rm sbo}^{}$. Consequently, longer observed timescales would generally arise as a result of the lower velocity and the correspondingly larger effective emission radius.



Although the timescale discrepancy could reflect an unrelated source, we discuss three potential distinct physical motivations for this difference, all of which have some pitfalls. An improved understanding of such cases may have to wait for additional examples:

\textbf{(i) CSM Asymmetry/Inhomogeneity:} 
While shock breakouts are often idealized as isotropic and homogeneous, spectro-polarimetric observations indicate that most, if not all, \iin\ CSM environments exhibit significant asymmetry \citep{Bilinski_2024}. Spectroscopic line profiles also reveal deviations from spherical symmetry along the line of sight \citep{Smith_2020_hcc}, typically arising from variations in the CSM's density distribution, optical depth, and radial gradients. Unlike in ordinary core-collapse \sne, the first light from \iin\ events often originate from the shock breakout through an optically thick CSM rather than directly from the stellar surface \citep{Calzavara_2004, Smith_McCray_2007, Smith_2017_book, Waxman_2017}.

Intrinsic anisotropies in the explosion itself can further amplify these effects; for example, simulations of RSG progenitors\footnote{Other progenitor types likely exhibit similar surface velocity differences due to variations in effective surface density and pressure. For example, 3D simulations of Type IIb progenitors show even greater surface inhomogeneity \citep{Goldberg_2025}.} show that the breakout velocity can vary by an order of magnitude across the stellar surface, reflecting differences in breakout radius and the local density structure near the photosphere \citep{Goldberg_2022_rsg, Goldberg_2022_sbo}. In \iin\ \sne, however, disentangling the geometry of the underlying ejecta from that of the more extended CSM interaction region requires additional spectroscopic or polarimetric constraints.

For asymmetric CSM configurations, the breakout duration is set by the time required for the shock to traverse the effective CSM surface along the observer's line of sight \citep{Couch_2011}. This differs from the cases of a stellar-surface breakout \citep{Falk_1978} or a spherically symmetric CSM \citep{Waxman_2017, Nagao_2020}. Consequently, the observed X-ray flash can vary substantially in both duration and luminosity depending on viewing angle, CSM extent, and optical depth.

Although asymmetric CSM structures are often invoked to explain extended breakout durations \citep{Couch_2011}, the same structure can theoretically shorten the duration. If the radiation from a shock propagating through an optically thick CSM encounters a locally thinner region with limited angular or physical extent, the effective shock-crossing time across that ``edge" of the CSM can be drastically reduced. In such cases, the X-ray flash may last only seconds -- set by the geometric light-crossing or local diffusion time -- rather than the typical minutes-to-hours expected for a spherical breakout \citep{Khatami_2024, Wasserman_2025_early}. Thus, a small patch of low optical depth can yield a brighter but much shorter burst, explaining the physical extra brightness (integrated burst energy of $E_{\gamma,\,\mathrm{iso}}^{}\approx 3\times10^{47}\;\rm erg$) and short duration of the \fermi\ lightcurve as photons escape more directly rather than diffusing through extended material \citep{Waxman_2017}.

As discussed in \S\ref{sec:model_discuss}, the long ($\sim$30$\; \rm day$) rise in the optical and UV light curves suggests that the shock breakout occurred within an optically thick CSM extending to roughly $10\;{\rm AU}$ (see \autoref{tab:mosfit_params}). In such an extended CSM, the initial radiation-driven shock is expected to evolve into a collisionless shock, producing strong X-ray emission over timescales comparable to the optical/UV rise \citep{Katz_2010, Wasserman_2025_wind}. These timescales are far longer than the brief $\sim$7$\,\rm s$ X-ray burst that was observed.

The short burst therefore likely originates from a much more compact region of the CSM, with a characteristic size of only a few tens of $R_\odot^{}$. Emission from such a region would naturally appear short in duration -- on the order of seconds -- because the light-travel time across it ($R/c$) limits how long the radiation can be observed. If the surrounding CSM is aspherical, this compact region may be only partially obscured, allowing a direct, transient view of the breakout from one localized patch rather than from the entire extended structure.

\rev{It is possible that multiple bound and unbound layers of CSM can create a combined wind and shell-like solution with $s\sim1.5$ \citep{Tsuna_2021}, which is within $2\sigma$ constraints of the modeled density profile exponent, $s$, from \S\ref{sec:model_discuss}. The shock breakout diffusion timescale as the forward shock propagates through multiple layers of CSM could potentially produce a shorter diffusion timescale. The difference in the CSM thickness and optical density between the bound and more compact inner regions to the outer layers may effect the observed timescale and observed flux density.}


\textbf{(ii) ``Choked" Jet:} 
A failed or ``choked" jet could have developed, quenched by the optically thick stellar envelope or less likely the dense CSM \citep{Piran_2019, Zegarelli_2024}. This scenario produces a mildly relativistic shock with a broad opening angle upon interaction with the CSM, capable of potentially reproducing the observed photon energies, the received flux density, and the burst duration \citep{Piran_2019}. However, one should always be cautious when interpreting any event as a jet, as geometrical degeneracies often allow jet models to fit the data even if it is not physically motivated. A shocked cocoon of a collapsar-driven jet choked in an extended CSM has been proposed as the solution for other X-ray flashes such as the Type Ib/c-BL, SN\,2025kg \citep{Srinivasaragavan_2025}, however, this event had a longer duration and a less dense CSM than \vxm, making them difficult to directly compare.

While such an interpretation remains plausible, most of the choked-jet theory has been developed for core-collapse \sne\ without considering CSM interaction \citep{Piran_2019, Zegarelli_2024}, and is only addressed in a few papers (e.g. \citealt{Duffell_2019}; \citealt{Srinivasaragavan_2025}), so further work will have to be done in understanding how the complexities of the environment/geometry has an impact on the photon energy spectrum and the duration.

Another caution, if the jet is `choked,' it is unlikely to produce the observed prompt emission. The high-quality \tess\ data suggest a weak or absent afterglow. While an afterglow without prompt emission is possible, generating prompt emission without an afterglow is difficult, as jet–medium interaction inevitably produces synchrotron radiation.

\textbf{(iii) Longer-Duration Event:} 
A third possible explanation is that the event lasted longer than observed, with only the peak detected because the low flux-density fell below the background noise. In this scenario, the total time-integrated energy from the interaction would be higher, and the true timescale would be more naturally explained by diffusion through a CSM. While this scenario can not be dismissed, the sharp peak in the \fermi\ light curve would still likely require both asymmetries in the CSM distribution and a clear line-of-sight/specialized observing angle to reproduce the observed features when compared to the softer peak in the two-dimensional \swift\ XRT simulations of \citet{Couch_2011}.

We are led to conclude that, so far, an inhomogeneous/aspherical CSM is likely the most physically plausible explanation. However, all three interpretations face theoretical challenges, which would require further observations or models.




\subsection{Lightcurve Modeling Discussion}\label{sec:model_discuss}

\begin{table*}
    \caption{Best-fit \mosfit\ model parameters for \vxm, compared with the median values and parameter distributions from a sample of 142 \iin\ \sne\ (R+V25) analyzed in \citet{Ransome_2025}. For each parameter, we report the value for \vxm, the sample median with uncertainties, and the difference between them in units of $\sigma$, accounting for the combined uncertainties. Parameters are grouped into ejecta, CSM, and other model components.}
    \label{tab:mosfit_params}
    \centering
    \begin{tabular}{ccccccc}
        \hline 
        \textbf{Ejecta/SN Params.} & & & & & & \\
        Sample &  & $M_{\rm ej}^{}$ & $v_{\rm ej}^{}$ & $T_{\rm min}^{}$ & $t_{\rm exp}^{}$ & --\\
        & & $M_{\odot}^{}$ & $\rm km\,s^{-1}_{}$ & $\rm K$ & $\rm days$ & --\\
        \hline 
        \vxm & & $38.8^{+6.6}_{-6.0}$ & $6918^{+161}_{-157}$ & $6310_{-144}^{+147}$ & $-0.86_{-0.09}^{+0.08}$ & \\
        \citet{Ransome_2025} Sample & & $20.1^{+19.0}_{-14.9}$ & $4721^{+3750}_{-2022}$ & $2587^{+4960}_{-2570}$ & $-10.4^{+5.7}_{-7.7}$ & --\\
        Difference &  & $0.94\sigma$ & $0.59\sigma$ & $0.75\sigma$ & $1.67\sigma$ & --\\
        \hline
        \hline
        \textbf{CSM Params.} & & & & & & \\
        Sample & & $M_{\rm CSM}^{}$ & $s$ & $\rho_0^{}$ & $R_0$ & $n$ \\
        & & $M_{\odot}^{}$ & -- & $10^{-12}\,\rm g\,cm^{-3}_{}$ & $\rm AU$ & --\\
        \hline
        \vxm & & $1.48^{+0.14}_{-0.13}$ & $1.40^{+0.08}_{-0.08}$ & $3.02^{+6.53}_{-1.92}$ & $12.3^{+14.0}_{-6.9}$ & $7.46^{+0.47}_{-0.29}$ \\
        \citet{Ransome_2025} Sample & & $1.26^{+6.68}_{-0.92}$ & $1.37^{+0.67}_{-0.93}$ & $6.31_{-5.81}^{+0.44}$ & $13.1^{+35.2}_{-9.7}$ & $9.44_{-1.80}^{+1.79}$\\
        Difference &  & $0.03\sigma$ & $0.04\sigma$ & $0.38\sigma$ & $0.05\sigma$ & $1.06\sigma$\\
        \hline
        \hline
        \textbf{Other Params.} & & & & & & \\
        Sample & & $N_{\rm H}^{}$ & $\sigma$ & $\delta$ & $\epsilon$ & --\\
        & & $10^{20}\,\rm cm^{-3}_{}$ & $\rm mag$ & -- & -- & --\\
        \hline
        \vxm & & $5.75^{+0.70}_{-0.63}$ & $0.126^{+0.006}_{-0.008}$ & $0$ & $0.5$ & -- \\
        \citet{Ransome_2025} Sample & & $0.08^{+6.23}_{-0.08}$ & $0.100^{+0.151}_{-0.005}$ & $0$ & $0.5$ & --\\
        Difference &  & $0.88\sigma$ & $0.17\sigma$ & -- & -- & -- \\
        \hline
        \hline 
    \end{tabular}
\end{table*}

The \mosfit\ analysis presented by \citet{Ransome_2025} involves 11 free parameters constrained by fitting, along with two fixed parameters. We compare the distribution of 142 \iin\ \sne\ from that study to the best-fit model parameters for \vxm, categorized into ejecta, CSM, and other components, as listed in \autoref{tab:mosfit_params}, including their $\sigma$-level tensions.

The ejecta-related parameters in the \mosfit\ CSM interaction model include: the ejected mass, $M_{\rm ej}^{}$; the characteristic velocity dispersion of the ejecta, $v_{\rm ej}^{}$; the minimum temperature, $T_{\rm min}^{}$, which reflects the model assumption that the photospheric radius expands until this temperature is reached before receding; and the explosion time, $t_{\rm exp}^{}$, relative to the first observation provided, which is $0.26\;{\rm days}$ before the first light is seen in \tess.

CSM parameters include: the CSM mass ($M_{\rm CSM}^{}$); the density profile exponent, $s$, where $\rho_{\rm CSM}^{} \propto R^{-s}$ and $R$ is the CSM radius; the density at the inner CSM radius ($\rho_0^{}$); and the $n$-index, which characterizes the geometric properties of the SN ejecta density profile.


Additional or fixed parameters include the host galaxy hydrogen column density, $N_{\rm H}^{}$; a white noise parameter, $\sigma$, accounting for systematic uncertainties; the outer ejecta density index, $\delta = 0$, following \citet{Ransome_2025}; and the energy conversion efficiency, $\epsilon=0.5$, for transforming kinetic energy into radiation during CSM interaction. A large hydrogen column depth could help explain the hardness ratio of the potentially related X-ray spectra as a the host galaxy will have greater absorption of lower-energy X-ray photons \citep{Couch_2011}.

The \mosfit\ lightcurve parameters infer a compact progenitor. The ejecta profile parameter, $n = 7 \to 10$ may refer to LBV or WR–like progenitors as opposed to RSG-like progenitors which are often fixed at $n = 12$ \citep{Colgate_1969, Matzner_1999, Kasen_2010, Chevalier_2011, Moriya_2013_bolo, Ransome_2025}. When coupled with the large total mass of the system $M_{\rm tot}^{}=M_{\rm ej}^{} + M_{\rm CSM}^{} = 40.3^{+6.7}_{-6.2}\;M_{\odot}^{}$, the high $T_{\rm min}^{}$, provides further evidence for a BSG/LBV-like progenitor. The stated total mass of the system does not include any prior mass loss by winds throughout the star's life during evolution, so the initial mass of the progenitor must have been even higher. The high ejecta mass and velocity produce an energetic explosion with approximately an order of magnitude more energy than typical core-collapse SN (e.g. \citet{Hamuy_2003}), with $E_{\rm KE}^{}=\frac{3}{10} M_{\rm ej}^{} v_{\rm ej}^{2} = \left(1.10^{+0.19}_{-0.18}\right)\times10^{52}\,{\rm erg}$. This high mass budget and high explosion energy of $\sim$10$^{52}$ erg is reminiscent of some other long-lasting SLSNe IIn like SN\,2003ma \citep{Rest_2011} and SN\,2015da \citep{Smith_2024_15da}.


The CSM density profile exponent, $s$ (some analyses such as \citet{Moriya_2013_bolo} refer to this parameter as $w$), is commonly fixed in many analyses to either $s=0$, representing the CSM as a dense optically thick shell, or $s=2$ representing a steady-wind like CSM distribution. Our best-fit value, $s=1.40^{+0.08}_{-0.08}$, has over a $5\sigma$ tension with either the dense-shell or the steady-wind scenario. Physically, the intermediate slope may reflect a complex mass-loss history involving multiple eruptive episodes superimposed on a steady wind, resulting in a structured CSM with underlying `clumpy', dense shells, which may result in the CSM distribution being inhomogeneous and asymmetric, or more generally, a time-variable mass-loss rate.

The assumption of spherical symmetry in \mosfit\ likely affects the inferred modeled parameters versus the `true' set of parameters \citep{Ercolino_2025}; due to the complex dynamics observed (see \citet{Smith_prep} for a more detailed analysis of the CSM dynamics) it is likely the environment is asymmetrically layered in both depth and density. As noted by \citet{Suzuki_2019}, various effects such as radiation hydrodynamics, Rayleigh-Taylor instabilities, Kelvin–Helmholtz instabilities, and Vishniac instabilities \citep{Vishniac_1983} along the effective ejecta-CSM surface will likely create deviations from the spherical symmetric models used, even when the distribution of both the ejecta and CSM are mostly spherical. However, the 200 day cut can affect observations for \iin\ \sne\ that rebrighten (e.g. \citealt{Ofek_2014_precursor, Smith_2009, Smith_2017_ip}), likely due to the ejecta interacting with multiple effective dense CSM layers or `clumps' \citep{Khatami_2024}. While recent work has been done in modeling these features by adding other effects such as a `breakout' parameter \citep{Khatami_2024}, the simple case CSM-interaction model in the \mosfit\ model can not account for such features, so complex lightcurve components that likely come from CSM layers are not reflected in the inferred parameters.

With asymmetric geometries, different regions/zones can be viewed simultaneously with different characteristic velocities \citep{Smith_2017_book}. At this level of complexity, the viewing angle and evolutionary stage at which the SN is observed play key roles in determining the luminous features seen in the light curve. We typically associate a highly non-spherical geometry with binary interaction, where the CSM has been `deflected' by another body, or where the mass loss is driven primarily by the binary interaction, as in an equatorially concentrated outflow made by Roche Lobe Overflow.

Another consideration is that \vxm\ is classified as an SLSN-IIn rather than a typical \iin\ SN, which while share the same general features as \iin\ \sne\ may affect the inferred parameter values \citep{Ransome_2025}. For SLSN-IIn events, \mosfit's CSM-interaction models can overestimate the CSM mass required to sustain long-duration light curves \citep{Suzuki_2021, Ransome_2025} in the absence of a potential central engine input. To mitigate this, we restrict our model fitting to within 200 days of the first observation, ensuring the data encompasses the entire rise and a portion of the decline to understand the fall-time.

\citet{Ransome_2025} compared the simplified CSM-interaction model in \mosfit\ to the numerical simulations of \citet{Dessart_2015} and found reasonable agreement in the inferred CSM masses. In our case, the model yields a physically plausible value of $M_{\rm CSM} = 1.48^{+0.14}_{-0.13}\;M_{\odot}^{}$. \\*[0.1cm]

\section{Conclusion}\label{sec:conclusion}

In this work, we present a multi-survey photometric analysis of the \iin\ \sne\ \vxm, incorporating data from \tess, \atlas, \ps, ZTF, \fermi, \swift, \gaia, and the LCOGT network. We model the high-cadence \tess\ rise, the first 200 days of the light curve using \mosfit\ (following \citet{Ransome_2025}), and the spectral energy distribution of the host galaxy, \host.

The early \tess\ light curve is well described by a broken power law with an index of $n=1.41\pm0.04$, significantly ($>$10$\sigma$) shallower than the canonical $\propto t^2$ rise typically assumed for early-time \sne\ emission. From this fit, we infer a time of first light, $t_0$, and find that the \mosfit-derived explosion time precedes the \tess\ rise by $0.26\;{\rm days}$, providing a strong temporal constraint and possible insight into the upper range of the shock breakout velocity distribution.

We assess the combined spatial and temporal coincidence likelihood between \vxm\ and the X-ray transient \rev{GRB\;191117A}, finding a $0.09\%$ spatial–temporal chance-coincidence probability, corresponding to a $3.3\sigma$ confidence that the events are associated. The brief duration of the X-ray event (on the order of seconds) potentially suggests interaction with a dense, possibly asymmetric CSM, the presence of a weak or `choked' jet, or a longer and fainter XRB event than observed. However, each consideration has issues that require further investigation.

The inferred CSM density exponent, $s=1.40^{+0.08}_{-0.08}$, indicates a steady pre-SN mass-loss history coupled to periods of eruptive or turbulent outflow, producing an inhomogeneous, clumpy CSM structure. Such an asymmetry may account for both the observed XRB lightcurve duration and the NIR--UV lightcurve properties.

Assuming \vxm\ and \rev{GRB\;191117A} are physically associated, the \rev{peak} photon energy \rev{of $E_{\rm peak}^{}=78.6^{+23.2}_{-19.4}\;{\rm keV}$} would typically indicate a more compact progenitor such as a LBV or WR star. However, in the presence of CSM, inverse comptonization and other effects could increase the average photon energy, and lower energy X-ray photons may get absorbed by the high hydrogen column-density, $N_{\rm H}^{}$, modelled in \mosfit. More strong evidence of \vxm\ having a compact progenitor comes from the \mosfit\ best-fit parameter for the geometric SN ejecta density, $n=7.46^{+0.47}_{-0.29}$, consistent with progenitors in the LBV--WR regime ($n\sim7$) and marginally extending to BSG-like structures ($n\sim7$–$10$). The total SN ejecta + CSM system mass, $M_{\rm tot}^{}=40.3^{+6.7}_{-6.2}\;M_\odot^{}$, provides corroborating evidence that \vxm\ has a compact progenitor; considering mass lost by winds during evolution, the initial mass would likely be well above 40 $M_{\odot}^{}$, whereas no known RSG progenitors are this massive. This is just within the upper range for compact progenitors, reinforcing the conclusion that \vxm\ originated from a massive, compact star embedded in a structured, asymmetric CSM.

If the association with \rev{GRB\;191117A} is real, \vxm\ represents one of the few known cases where a superluminous \iin\ has been associated with an X-ray burst at shock breakout. The high-cadence data from \tess\ alongside the \fermi\ X-ray burst offers an unprecedented window into the onset of core collapse and shock breakout in a structured CSM.

\section*{acknowledgments}

\rev{We would like to thank the anonymous reviewer for the helpful feedback and notes to improve the quality of the research.}
We greatly appreciate the help of Jan Eldridge for help with initial interpretation and early \rev{modeling}. We also want to acknowledge the engaging discussions with Ryan~J.~Foley, Ore Gottlieb, Gautham Narayan, Justin Pierel, Tyler Pritchard, Melissa Shahbandeh, and Louis-Gregory Strolger that helped frame aspects of the analysis and provided stimulating discussions. We acknowledge access to the few Pan-STARRS data points from the Pan-STARRS NEO survey data.

\rev{We would also like to thank Rachel Hamburg and Eric Burns for reaching out with improvements for the \fermi\ analysis while under review, their help and advice strengthened the final results and analysis.}

Z.G.L., R.R.H., C.M. and P.M. are supported by the Marsden Fund administered by the Royal Society of New Zealand, Te Ap\={a}rangi under grants M1255 and M1271. Z.G.L. would like to acknowledge the support of the School of Physical and Chemical Sciences Travel Grant, which contributed to the success of this research. R.R.H. is also supported by the Rutherford Foundation Postdoctoral Fellowship RFT-UOC2203-PD. H.R. is supported by an Australian Government Research Training Program (RTP) Scholarship. Q.W. is supported by the Sagol Weizmann-MIT Bridge Program. S.J.S. acknowledge funding from STFC Grant ST/Y001605/1, a Royal Society Research Professorship and the Hintze Family Charitable Foundation. The Flatiron Institute is supported by the Simons Foundation. \rev{K.L.\ and M.L.\ was supported by the `SeismoLab' KKP-137523 \'Elvonal grant of the Hungarian Research, Development and Innovation Office (NKFIH), and by the LP2025-14/2025 Lendület grant of the Hungarian Academy of Sciences. K.L.\ thanks the financial support provided by the undergraduate research assistant program of Konkoly Observatory.}

This work was partially supported by NASA ADAP grant 80NSSC22K0494.

\rev{The material is based upon work supported by NASA under award number 80GSFC24M0006.}

This paper includes data collected with the TESS mission, obtained from the MAST data archive at the Space Telescope Science Institute (STScI). Funding for the TESS mission is provided by the NASA Explorer Program. STScI is operated by the Association of Universities for Research in Astronomy, Inc., under NASA contract NAS 5–26555.

This work has made use of data from the Asteroid Terrestrial-impact Last Alert System (ATLAS) project. The Asteroid Terrestrial-impact Last Alert System (ATLAS) project is primarily funded to search for near earth asteroids through NASA grants NN12AR55G, 80NSSC18K0284, and 80NSSC18K1575; byproducts of the NEO search include images and catalogs from the survey area. This work was partially funded by Kepler/K2 grant J1944/80NSSC19K0112 and HST GO-15889, and STFC grants ST/T000198/1 and ST/S006109/1. The ATLAS science products have been made possible through the contributions of the University of Hawaii Institute for Astronomy, the Queen’s University Belfast, the Space Telescope Science Institute, the South African Astronomical Observatory, and The Millennium Institute of Astrophysics (MAS), Chile.

The Pan-STARRS1 Surveys (PS1) and the PS1 public science archive have been made possible through contributions by the Institute for Astronomy, the University of Hawaii, the Pan-STARRS Project Office, the Max-Planck Society and its participating institutes, the Max Planck Institute for Astronomy, Heidelberg and the Max Planck Institute for Extraterrestrial Physics, Garching, The Johns Hopkins University, Durham University, the University of Edinburgh, the Queen's University Belfast, the Harvard-Smithsonian Center for Astrophysics, the Las Cumbres Observatory Global Telescope Network Incorporated, the National Central University of Taiwan, the Space Telescope Science Institute, the National Aeronautics and Space Administration under Grant No. NNX08AR22G issued through the Planetary Science Division of the NASA Science Mission Directorate, the National Science Foundation Grant No. AST–1238877, the University of Maryland, Eotvos Lorand University (ELTE), the Los Alamos National Laboratory, and the Gordon and Betty Moore Foundation. Pan-STARRS is now primarily funded to search for near-earth asteroids through NASA grants NNX08AR22G and NNX14AM74G. The Pan-STARRS science products for LIGO--Virgo--KAGRA follow-up are made possible through the contributions of the University of Hawaii's Institute for Astronomy and Queen's University Belfast and the University of Oxford. 

This work is based on observations obtained with the Samuel Oschin Telescope 48-inch and the 60-inch Telescope at the Palomar Observatory as part of the Zwicky Transient Facility project. ZTF is supported by the National Science Foundation under Grants No. AST-1440341 and AST-2034437 and a collaboration including current partners Caltech, IPAC, the Oskar Klein Center at Stockholm University, the University of Maryland, University of California, Berkeley , the University of Wisconsin at Milwaukee, University of Warwick, Ruhr University, Cornell University, Northwestern University and Drexel University. Operations are conducted by COO, IPAC, and UW.

\rev{The operation of the Konkoly RC80 telescope was
supported by the GINOP 2.3.2-15-2016-00033 project of
the National Research, Development and Innovation Office (NKFIH), Hungary, based on funding from the European Union.}

This research is based on observations made with the Neil Gehrels Swift Observatory, obtained from the MAST data archive at the Space Telescope Science Institute, which is operated by the Association of Universities for Research in Astronomy, Inc., under NASA contract NAS 5–26555.

This research has made use of data and/or software provided by the High Energy Astrophysics Science Archive Research Center (HEASARC), which is a service of the Astrophysics Science Division at NASA/GSFC.

The Fermi Gamma-ray Data Tools are partially funded through the NASA ADAP Grant 80NSSC21K0651 and the NASA SMD Open Source Tools, Frameworks, and Libraries Grant 80NSSC22K1741.

We acknowledge ESA Gaia, DPAC and the Photometric Science Alerts Team (\url{http://gsaweb.ast.cam.ac.uk/alerts}).


This work makes use of observations made with the Sinistro instrumentation at McDonald Observatory, operated by the Las Cumbres Observatory.

\rev{This work makes use of observations from the Las Cumbres Observatory network.}




\section*{Data Availability}

All of the data and codes are made public and available at \url{https://github.com/ZacharyLane1204/SN2019vxm}.

\software{
\texttt{AstroColour} \citep{Lane_2025_colour}, 
\texttt{ATClean} \citep{Rest_2025},
\texttt{dynesty} \citep{Higson_2019, Speagle_2020},
\texttt{FrankenBlast} \citep{Nugent_2025},
\texttt{gdt-fermi} \citep{Goldstein_2023},
\mosfit\ \citep{Guillochon_2018},
\texttt{pysynphot} \citep{STScI_2013},
\texttt{TESSELLATE} \citep{Roxburgh_2025},
\texttt{TESSreduce} \citep{Ridden_2021},
\texttt{astropy} \citep{Astropy_2013, Astropy_2018, Astropy_2022}, 
\texttt{astroquery} \citep{Ginsburg_2019},
\texttt{healpy} \citep{Gorski_2005, Zonca_2019}, 
\texttt{matplotlib} \citep{Hunter_2007}, 
\texttt{numpy} \citep{Harris_2020}, 
\texttt{pandas} \citep{McKinney_2010},
\texttt{scipy} \citep{Virtanen_2020}
}

\bibliography{bibliography}{}
\bibliographystyle{aasjournal}

\appendix
\renewcommand{\thefigure}{A\arabic{figure}}
\setcounter{figure}{0}

\section{\mosfit\ Corner Plot}

\begin{figure*}[h]
    \centering
    \includegraphics[width=1\linewidth]{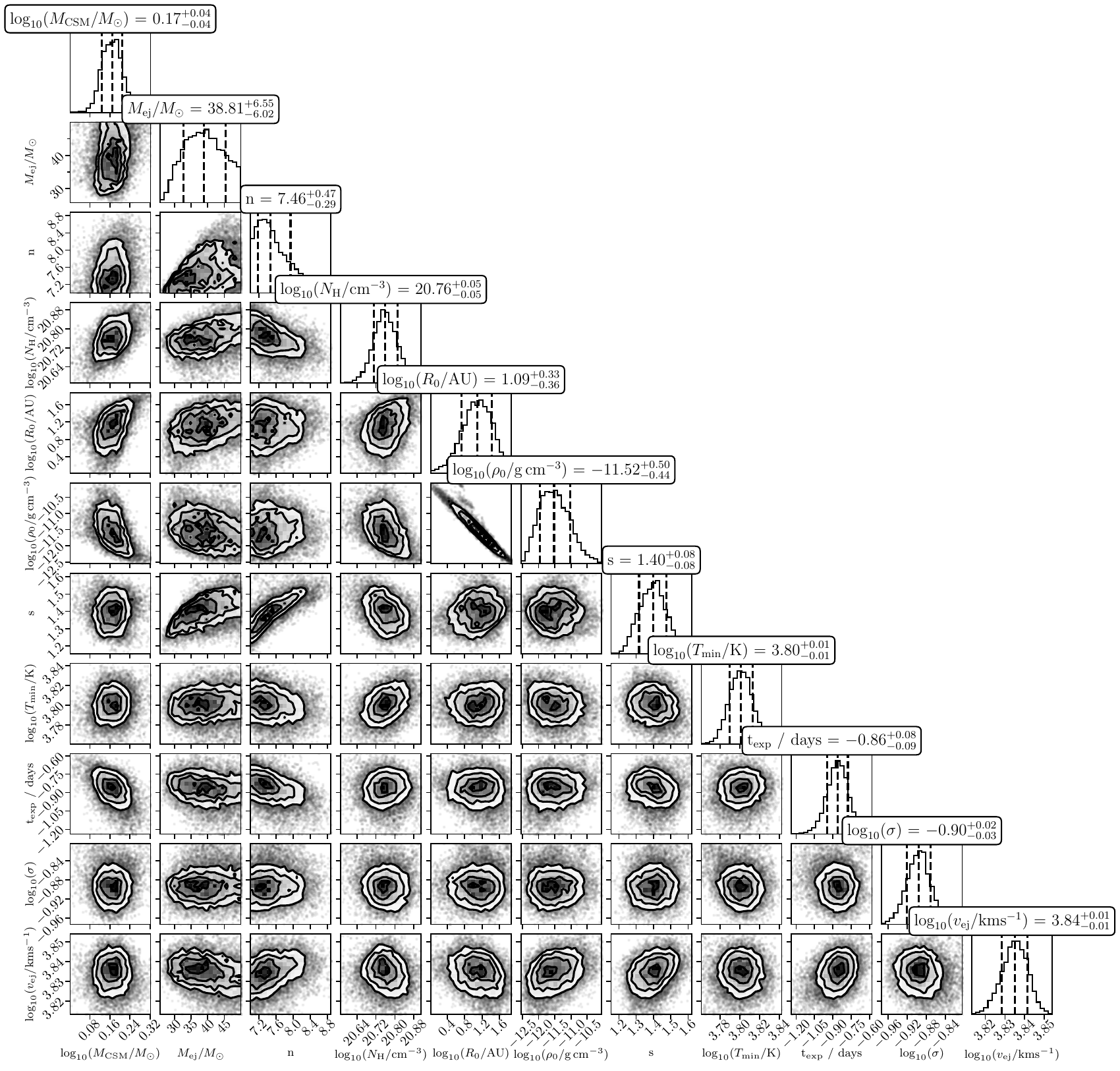}
    \caption{The corner plot for the CSM-interaction \mosfit\ lightcurve parameters showing their posterior distribution for \vxm. The diagonal panels show the marginalized probability distributions for each parameter, while the off-diagonal panels illustrate their covariances and parameter degeneracies. We have marked the weighted median as well as the 16\% and 84\% quantiles of the posterior distribution with dashed lines.}
    \label{fig:mosfit_corner}
\end{figure*}

\end{document}